# LECTURE II: COMMUNICATIVE JUSTICE AND THE DISTRIBUTION OF ATTENTION

Seth Lazar

Machine Intelligence and Normative Theory Lab

Australian National University[*]

## ABSTRACT

Algorithmic intermediaries govern the digital public sphere through their architectures, amplification algorithms, and moderation practices. In doing so, they shape public communication and distribute attention in ways that were previously infeasible with such subtlety, speed and scale. From misinformation and affective polarisation to hate speech and radicalisation, the many pathologies of the digital public sphere attest that they could do so better. But what ideals should they aim at? Political philosophy should be able to help, but existing theories typically assume that a healthy public sphere will spontaneously emerge if only we get the boundaries of free expression right. They offer little guidance on how to intentionally constitute the digital public sphere. In addition to these theories focused on expression, we need a further theory of communicative justice, targeted specifically at the algorithmic intermediaries that shape communication and distribute attention. This lecture argues that political philosophy urgently owes an account of how to govern communication in the digital public sphere, and introduces and defends a democratic egalitarian theory of communicative justice.

[*] This essay, based on my 2023 Tanner Lecture on AI and Human Values at Stanford, will be published in my forthcoming book, *Connected by Code: How AI Structures, and Governs, the Ways we Relate*, with Oxford University Press, alongside commentaries by Joshua Cohen, Marion Fourcade, Renée Jorgensen and Arvind Narayanan, and my response to those commentaries. Please refer where possible to the published version. See the end for acknowledgments.

*The ties which hold [people] together in action are numerous, tough and subtle. But they are invisible and intangible. We have the physical tools of communication as never before. The thoughts and aspirations congruous with them are not communicated, and hence are not common. Without such communication the public will remain shadowy and formless, seeking spasmodically for itself, but seizing and holding its shadow rather than its substance. Till the Great Society is converted into a Great Community, the Public will remain in eclipse. Communication can alone create a Great Community. (*[Dewey, 2016 (1926): 170.](Dewey, 2016 (1926): 170)*)*

## 1. INTRODUCTION

In 'Governing the Algorithmic City', I introduced a general model of how algorithmic intermediaries mediate our social relations, and in doing so enable and exercise governing power, which urgently demands justification or elimination. I then showed how algorithmic governance poses pressing questions of political authority, procedural legitimacy, and substantive justification, our theories of which have been developed for a quite different socio-political context, where extrinsic control of social relations by coercive laws was prevalent. In this Lecture, I apply this theoretical approach to one aspect of the Algorithmic City—the digital public sphere—broaching an urgent philosophical question: how should liberal egalitarian democracies shape public communication and distribute collective attention?

Concern about the health of the digital public sphere can sometimes seem ahistorical, forgetting how collective communication has always fallen short of our ideals.[2] And it can be overstated, ignoring important successes like how the internet has enabled structurally disadvantaged communities to come together and forge a common identity.[3] But however measured one's approach, the pathologies of online communication are hard to ignore.

In Lecture I, I argued that algorithmic power is presumptively morally suspect, but can be justified when used for the right ends, in the right ways, by those with the proper authority to do so. Algorithmic governance of the digital public sphere falls short on each count.[4] What is power used for? Mostly the (sometimes forlorn) pursuit of returns on investment to the private owners of these platforms. This typically means maximising time on platform to support surveillance- and engagement-based advertising.[5] Who exercises power? Sometimes conflicted and capricious platform owners;[6] or else harassed and precarious click-workers;[7] or—at best—advisory councils and oversight boards that lack any democratic authority.[8] How is power exercised? Privately, with little to no community authorship, no expectation of consistency in application, and little accountability for bad judgements.[9] What is the net result? Our digital public sphere, which in principle affords an unprecedented opportunity to create Dewey's 'Great Community', is instead host to numerous pathologies—from the pollution of our information environment, to individual and collective practices of silencing and abuse, to targeted and stochastic manipulation. We could clearly be doing better.

But what would better look like? What should we aim for, beyond just trying to put out each spot fire as it flares up? Political philosophy could help us answer these questions. But it has engaged too little with the fast-changing realities of the digital public sphere.[10] And its prevailing framework for evaluating public communication is maladapted to the task ahead. This Lecture tries to make progress. I argue that the digital public sphere's shortcomings require us to rethink how online platforms *shape public communication* and *distribute attention*. We need normative principles to guide those tasks. Philosophers are likely to reach for the toolkit of freedom of expression. Building on Dewey's argument for the crucial role of communication in creating a 'Great Community', I will argue that we instead must craft a new account of *communicative justice*,[11] which explicitly aims to guide the intentional constitution of a healthy digital public sphere.[12] I then show how this ideal can offer guidance on substantive justification, proper authority, and procedural legitimacy in the governance of the digital public sphere.

## 2. SHAPING COMMUNICATION, DISTRIBUTING ATTENTION

Let's start with some stipulative definitions. The public sphere is the environment for public communication. *Communication* is the social practice whereby one party expresses themselves, and another attends to that expression. *Expression* involves the creation of meaningful content through words, images, or other artefacts. *Attention* involves orienting one's mind towards that content, and to a greater or lesser degree processing it. Attention can be fleeting or deeply engaged. Communication is *public*, let's assume, when those communicating lack a reasonable expectation of privacy. Since our environment for public communication is now digital, the digital public sphere just is the public sphere. Some theorists reserve the idea of the public sphere for the domain of public communication that focuses on politics.[13] I think this is conceptually and normatively untenable—not only is everything political, but every medium for public communication will eventually be used for political discussion (as traditionally understood).[14] Many of the challenges of public communication would be easier to address if we had a public sphere that could easily be

---

partitioned into different categories of communication.[15] Unfortunately, the actual public sphere is not like this.

The shortcomings of the digital public sphere mostly fall into one of three rough categories: abuse; epistemic pollution; and manipulation. The first encompasses direct and indirect abuse, harassment,[16] and silencing.[17] It involves using online communication to harm others directly (whether individually, or as one element in a collective harm).[18] The second gathers together communicative practices that make it harder for us to form accurate beliefs. This includes the production and circulation of misleading online content through conspiracy theories, misinformation and disinformation,[19] coordinated inauthentic behaviour,[20] obfuscation and 'flooding'.[21] Manipulation can be variously defined. For my purposes, I'll use it to mean the practice of using communication to influence others to make decisions in ways that compromise their autonomous agency.[22] This includes both intentional 1:1 manipulation, for example, where one party tries to radicalise another,[23] and perhaps-unintended manipulation of whole populations, as when platforms contribute to affective polarisation,[24] or to the rise of pathological body-image anxiety.[25] These categories, of course, are not mutually exclusive (a single communicative act could instantiate all three).

The digital public sphere is largely constituted by algorithmic intermediaries — computational systems that mediate social relations, in this case by enabling mediatees to communicate. In the near future, Language Model Agents acting as universal intermediaries might become central to the public sphere (I speculate about one possibility in Section 5D below). For now, however, LLMs are just one element in the algorithmic toolkit of the large platforms that have mediated billions of people's online lives for the past decade and more. I am thinking in particular of social media sites such as TikTok, Facebook, Instagram, Twitter, Discord, Reddit, SnapChat, YouTube, Mastodon and others, as well as search sites such as Google, and I suppose one must now say Bing.[26] My argument in this lecture rests on one or both of two conjectures about the relationship between those platforms and the pathologies of online communication. First: the design of online platforms is at least part cause of the shortcomings of the digital public sphere. Second: even if the first conjecture is in doubt, platform design is an important lever for building a healthier public sphere.

---

To motivate these two conjectures, we must start by identifying the specific features of platform design that are likely to contribute to the shortcomings of the digital public sphere, and that conversely might be (part of) its salvation. I think we can roughly parse the design of online platforms into two broad functions: shaping communication and distributing attention.

By *shaping communication*, I mean dictating the conditions under which people can communicate online.[27] This starts with *Identity Management*, determining whether and how a user's identity is verified to the platform and to other users.[28] Does the platform require real names or permit anonymity? Are identities verified or not? What options for account management and presentation are enabled? Next it concerns *Content Creation and Sharing*, which dictates the formats in which users may communicate with one another—for example short or long-form text, image, audio and video; as well as whether users can edit or co-author posts, and whether posts are temporary or permanent.[29] It concerns too the control users have over their audience—the degree to which they can restrict the distribution of their posts. *Interaction and Feedback* then covers how others can respond to that communication—what non-linguistic (e.g. likes, downvotes, upvotes) and linguistic options do they have to engage? Are quote-posts enabled, or threaded conversations? Are private messages enabled and if so are they encrypted or not?

Let's call the foregoing elements of platform design the platform's *architecture*. Platform architecture dictates your options for communication in the digital public sphere. But 'pre-emptive governance' alone is insufficient to prevent undesirable communication online: platforms also need *Safety and Moderation* practices to police harmful or otherwise undesirable behaviour.[30] Can users report other users for harmful communication? What happens if they do? What are the limits of permissible communication on the platform, and what are the consequences when those limits are ignored? For years, platforms desperate to preserve a veneer of neutrality (and perhaps an early ethos of pro-social community) sought either to avoid moderation, or to conceal its practice.[31] This veil has long-since dropped, and a full range of moderation practices are being pursued, from automated tools relying on Large Language Models, to vast armies of exploited click-workers, to ambitious attempts to invent quasi-judicial advisory councils.[32]

Platforms dictate how we may communicate online. Importantly, this involves not only the disciplinary power of post facto enforcement, but also the productive power of pre-emptive governance, determining the formats, registers, and audiences with which we can communicate.[33] But they do more than this. Through their *Discovery and Curation* practices, platforms also substantially shape the

---

*distribution of attention* online. Attention is the process of cognitive engagement that enables a hearer to receive a message expressed by a speaker. As has long been remarked, when expression is so cheap as to be almost costless, attention becomes the scarce resource.[34] Platforms therefore devote extraordinary efforts to discovery and curation practices that efficiently allocate attention.[35] While attention is fundamentally a process in the mind of an individual (the one who attends), we can also speak of *collective* attention being distributed by platforms, as they direct many individuals to attend to the same thing, and then facilitate the shared experience of that common attention through enabling interaction, commentary, and other forms of participatory engagement.[36]

The process of allocating (individual and collective) attention is often called 'algorithmic amplification'.[37] This evocative phrase can be misleading in two distinct ways. First, it risks reifying 'the algorithm'. Amplification algorithms—recommender systems—are always part of broader sociotechnical systems. They rely on signals from the platform architecture (for example, tracking a user's previous engagements to predict future ones).[38] Platforms also enable non-algorithmic amplification through their design and user-controlled discovery and curation—for example, when they rely on individuals subscribing to or following other accounts.[39] Adopting the 'network' model, whereby users can find content and other users based on their own network, also contributes to amplifying some voices above others; so does enabling muting and blocking. It is also debatable whether a CEO instructing his engineers to boost the voice of some specific users really counts as *algorithmic* amplification.

In addition, second, the very concept of amplification is hard to pin down, because it presupposes a baseline against which it can be measured.[40] What would that baseline be? If a platform serves any function at all, then aren't *all* posts on that platform being amplified, relative to their distribution if hosted on a private server? Or is a post amplified just in case platform choices lead to it being viewed more than the average post? More than it would have been viewed in the absence of those choices? Or more than it would have been viewed in the absence of any kind of platform intervention? Can we even make sense of this last idea?

In one sense, *everything* platforms do potentially amplifies the voices of those who contribute to that platform, just in virtue of their coalescing an audience. In a similar way, if you put up a noticeboard outside a village store, on which people can place notices, then the mere fact of putting up that noticeboard amplifies those voices relative to the alternative where no board exists. We can, however, distinguish between the passive amplification of coalescing an audience, and the

---

active amplification of distributing content to members of that audience.

Active amplification presupposes distribution: platforms amplify posts when they distribute them to more users, in ways that make them more visible.[41] This includes both responding to search queries, and, most paradigmatically, making recommendations. The basic function of recommender systems is twofold: to filter and to rank. The universe of possible content to which you could be exposed online is vast. Filtering is the process of reducing that to a manageable shortlist.[42] Ranking orders the list. Amplification can be defined as a function of two things: the degree to which a given piece of content, X, has been included on those shortlists, as a candidate for distribution, and the weights it has been given in the rankings on those lists. On this approach, for any X, at any given time, there is a simple scale for amplification ranging from zero to one. The zero point is when X is excluded from all shortlists, and left as a bare post wherever it is originally posted. Total amplification is achieved when X is included and ranked on top of all shortlists (as seems to have happened when the CEO of X, fka Twitter, complained about not getting enough attention on the platform). Algorithmic amplification is the subset of amplification carried out by recommender systems. Note that platforms can amplify not only posts, but also voices and communities, by recommending accounts to follow, or groups to join.

Thinking in terms of filtering and ranking also helps us to understand the complement of amplification, sometimes called demotion, reduction, deboosting, or deamplification.[43] Filtering is a process of exclusion as well as inclusion. It determines all the content that you *won't* see. For example, a post can be considered as a candidate for everyone's feed, feeds of a superset of followers short of everyone, feeds of followers only, a subset of followers, or only those who view the user's profile. Reduction or deamplification is essentially the process of moving along this list.

Tarleton Gillespie regards this kind of filtering as a species of content moderation, since both approaches restrict the visibility of what we communicate online.[44] Alternatively, one could argue that content moderation itself is just part of the curation process, another method for distributing attention. Alternatively again, one could say that everything—including curation—is content moderation.[45] While there are clearly some fuzzy boundaries here, and these are all reasonable interpretations of these terms, I think some conceptual clarification could be helpful. The concept of content *moderation* will, for my purposes, be defined by the reasons on which it is based, not by the form it takes or the outcome to which it leads. Moderation is action taken by a platform to enforce its rules. Action counts as enforcement only if it is directed at a party that is held to have transgressed those rules. Moderation, then, includes not just removal of content, but also suspension and banning of accounts. And it could include filtering of user content, if it is being filtered on grounds that the user has breached platform policies. This would be filtering as moderation. But when posts are filtered on grounds that they are undesirable, but do not breach any rules and so are not objects of enforcement of those rules, this is filtering as curation, not moderation.

---

[41] From here on, when I refer to amplification simpliciter I mean active amplification.
[42] Bengani *et al.*, 2022
[43] Ananny, 2019; Keller, 2021a; Gillespie, 2022; Narayanan, 2023
[44] Gillespie, 2022.
[45] This is basically the view of Grimmelmann, 2015, but has been recently articulated in this form by Evelyn Douek.

Platforms shape communication and distribute attention. They do not dictate precisely how we may communicate with one another online.[46] They do not unilaterally decide what we will attend to. But their design choices make some communicative practices easy and others hard. They impose obstacles and remove constraints.[47] They discourage and they incentivise.[48] They exercise — at extraordinary scale — intermediary power, constitutively shaping the social relations that they mediate.

A large body of research suggests that platforms' design choices causally contribute to epistemic pollution, abuse, and manipulation.[49] For example, as well as distributing attention, platforms aim to maximise the amount of attention available for distribution; platform design encourages people to remain on the platform for longer, curation practices display content that people are more likely to attend to and engage with.[50] Inflammatory or otherwise emotive content tends to attract attention, so communication is shaped and attention distributed in ways that foster volatility.[51] This incentivises deception, hostility and manipulation. And it applies with even greater force to businesses (news and otherwise) whose viability depends on their ability to engage audiences on social media.[52] More generally, platforms provide would-be manipulators with unparalleled resources with which to reach out directly to their targets — and to monetise their credulity.[53]

---

[46] Many scholars have shown how users push back against these attempts at control, either through exit or through co-opting platform affordances to their advantage. E.g. Gillespie, 2018: 23, but also Van Dijck, 2013; Bucher, 2018.

[47] This reaches its apotheosis in the cultural platforms that control large sections of the arts (Spotify, Apple Music, Audible, Kindle). These platforms present themselves as intermediaries between consumers and the art that they love, but they are just as much intermediaries between artists and their audience, and exercise unilateral control over the chokepoints they have created between expression and consumption. See Giblin and Doctorow, 2022.

[48] These are the 'affordances' of platform architecture — how the features of a technology affect its functions, in particular by making certain kinds of uses and outcomes more or less likely, for example by inviting and incentivising some uses, and discouraging or frustrating others. See Gibson, 1966. For a comprehensive recent theory of affordances, see Davis, 2020. Thanks to Jenny Davis for discussion of this point.

[49] Some relevant highlights: Andrejevic, 2013; Burgess et al., 2017; Settle, 2018; Vaidhyanathan, 2018; Thorburn et al., 2023. On polarisation specifically see Kubin and Von Sikorski, 2021.

[50] For excellent and balanced overviews of this vast literature, see Bengani *et al.*, 2022; Narayanan, 2023.

[51] E.g. Crockett, 2017; Settle, 2018; Vaidhyanathan, 2018; Andrejevic, 2020; Carpenter et al., 2020; Munger and Phillips, 2020; Brady et al., 2022.

[52] Keller, 2021a: 33; Diakopoulos, 2019. This is an obvious reason why Benkler et al., 2018 miss the mark. They argue that social media are not responsible for the state of democracy because, for independent reasons, right-wing media have cleaved from the rest of the media ecosystem, and circulate and legitimate misinformation and radicalising content. But they don't adequately consider how incentives created by social media have changed that ecosystem — e.g. the success of competitors like Breitbart on Facebook clearly led Fox to compete on the same terms to protect their position at the pinnacle of the right-wing media ecosystem.

[53] Munger and Phillips, 2020.

They knit discrete communications by many individuals into a single collective harm (for example by identifying topics that are trending, or through likes or upvotes for abusive or otherwise harmful statements).[54] They give us insight into the political views of our weak ties, exacerbating in-group and out-group dynamics, contributing to affective polarisation (a species of manipulation in my usage).[55]

While there is a long litany of causal pathways by which platforms' choices about how to shape communication and distribute attention contribute to each of these pathologies, there is also some dissent. Some think that the pathological state of our communications ecosystem is overstated; others think that things are bad, but the fault lies with us, not with the platforms.[56] I want to mostly sidestep these empirical debates.[57] Even the most ardent boosters of the digital public sphere could hardly deny that *we could be doing better*.[58] And even if the shortcomings of online communication are ultimately due more to *our* pathologies as users of online platforms than to the platforms' decisions, how we shape communication and distribute attention are clearly essential levers in rebuilding the digital public sphere. Just asking people to be nicer and more truthful online is obviously not going to cut it. Even government regulation must be implemented and enforced by online platforms; and most online pathologies are due not to obviously regulatable behaviour by individuals—individual acts that are sufficiently serious to warrant some kind of legal enforcement—but rather to emergent effects of mass communication, such that no individual is plausibly held accountable for the resulting harms.[59]

---

[54] Barnes, 2022; Brock, 2022.

[55] Settle, 2018.

[56] I have in mind some of the chapters in Persily and Tucker, 2020, in particular Barberá, 2020, as well as (especially) Benkler *et al.*, 2018. And for a canonical review of sources, see Clegg, 2021. For a very useful response to this kind of quietism, see Andrejevic, 2020; Andrejevic and Volcic, 2021.

[57] Due to influential speculation by Cass Sunstein and Eli Pariser, much of the empirical debate has focused on whether online platforms contribute to the creation of echo chambers (homophilous communities who reinforce one another's views), or filter bubbles (individual experiences of the curated internet that have the same effect). In general, neither phenomenon seems borne out by the empirical evidence. But the unwarranted conclusions are often drawn from debunking these particular causal pathways to a pathological digital public sphere: if we don't live in echo chambers or filter bubbles, then our online communication environment is in rude health. This is wrong. If online communication makes it hard to form accurate beliefs, fosters abuse in all its forms, and supports both agential and environmental manipulation, then that's a problem whether or not it is explained by echo chambers and filter bubbles. Relatedly, I think that investigations of the health of the digital public sphere that focus on analysing the distribution of content across different online accounts, focusing either on selective content exposure, or the prevalence of harmful content, are often misleading because of lack of access to relevant data, very narrow methodologies, and because of their failure to identify the communicative context for any given piece of content. In my view we learn more from methods that either combine quantitative studies with qualitative research, or focus on qualitative research that shows the specific psychological pathways by which online communication undermines our epistemic practices, or facilitates abuse and manipulation.

[58] For a similar approach to mine here, see Cohen and Fung, 2021.

[59] For example, the Digital Services Act regulates platforms by holding them accountable for

Rather than revisit epidemiological debates about the causes of our communicative dysfunction, I want to ask instead where we go from here.[60] We should start by recognising that algorithmic intermediaries in general, and online platforms in particular, *govern* the digital public sphere. To govern, as I claimed in Lecture I, is to make, implement, and enforce the constitutive norms of an institution or community. Through platform design, including architecture, moderation, and curation practices, platforms determine the norms of online communication, implement them, and enforce them. They thereby constitute the social relations that they mediate; through both intermediary and extrinsic power they shape power relations among the governed, and shape the shared terms of their social existence—over time reshaping social structures at large. Platforms are responsible for enabling some people to limit others' liberty, and to have power over others in ways that undermine relational equality. And they govern collective communicative practices that the members of those practices have presumptive rights to shape.

Even if the pathologies of the digital public sphere are in large part ultimately due to us, the users, algorithmic intermediaries must determine how to govern our online interactions better, or else be implicated in the deception, harassment, and manipulation that they help to constitute. The pathologies of human social interaction *in general* are due to us, the people doing the interacting. Governments do not escape the obligation to protect values such as individual freedom, relational equality, and collective self-determination simply on grounds that the threats to those values are *not their fault*, but the fault of those who would oppress, manipulate and deceive. Of course, it's an open question whether online platforms should have ultimate authority to govern the digital public sphere—I return to this in Section 5—but given that they alone are in a position at least to implement and enforce norms of online communication, they clearly have some role to play, even if only as the enactors of democratically mandated laws.[61]

And irrespective of whether platforms or states are ultimately responsible for governing the digital public sphere, we may surely conclude that how we shape communication and distribute attention will significantly affect the health of the digital public sphere.[62] At the very least, we know for sure that shaping communication and distributing attention in such a way as to maximise people's online engagement, increase the amount of attention there is to distribute, and so maximise advertising revenue is not the way to go. We are therefore forced to ask: what should we aim at instead?

On one approach, we hold the basic function of our communicative ecosystem

---

systemic risks that they cause, and thereby incentivising them to govern the users of platforms in ways that reduce systemic risks.

[60] On the epidemiological point, see Cohen and Fung, 2021; Reich *et al.*, 2021.

[61] The question of *which democracy* is of course a persistent thorn in the side of a democratically-inspired approach to platform governance.

[62] To be very clear: while online platforms such as social media companies currently shape communication and distribute attention, other algorithmic intermediaries may take their place. In particular, Language Model Agents based on Large Language Models may come to mediate our access to information and communication in even more comprehensive ways than do existing online platforms. Nothing in this paper depends on the fact that the specific algorithmic intermediaries in question are online platforms, rather than some subsequent counterpart that plays the same role.

constant, and then aim to identify each particular pathology and develop countermeasures that mitigate its effects. We acknowledge that online platforms are fundamentally profit-maximising corporations, but argue that their pursuit of profit must be constrained by the need to implement and improve these countermeasures.

This 'whac-a-mole' strategy is obviously important, but I think we need something more: we need a positive ideal to aim at when shaping communication and distributing attention. This is, first, for the obvious reason that existing pathologies might be reliable side-effects of pursuing our goals as we currently understand them (roughly, the pursuit of profit by online platforms). So if we don't rethink our goals, we'll be fixing pathologies with one hand while causing them with the other.

Second, we can't address some pathologies without articulating a positive goal—for example, if recommender systems aren't going to optimise for engagement, what should they optimise for? We need positive ideals in order to say.

Third, sometimes our countermeasures to pathologies will conflict with each other. Understanding our positive goals will afford a common currency with which to manage these trade-offs.

Lastly, in the words of just war theorist Paul Ramsay, 'what justifies, limits'.[63] Understanding what we are aiming for will help us understand constraints on the means that can permissibly be used to get there. In particular, the justification of governing power depends on it being used for the right ends, according to the right procedures, by those with proper authority to do so (I call these the 'what', 'who', and 'how' questions).[64] Understanding what those who shape communication and distribute attention should aim at will help us understand what procedures they should follow, and who has the right to exercise that kind of power.

## 3.   FROM FREEDOM OF EXPRESSION TO COMMUNICATIVE JUSTICE

Freedom of expression is an essential component in the political philosophy of communication, but it must take its place in a broader account of *communicative justice* to guide us in shaping communication and distributing attention.

Start by distinguishing (crudely), between theories of negative and positive freedom of expression. Negative freedom of expression (often associated with US first amendment jurisprudence) aims to vindicate individuals' negative rights to self-expression against interference, especially by the state.[65] Positive freedom of expression (often more associated with European jurisprudence) sees speakers' interests in self-expression as just one set of interests, alongside the interests of audiences and bystanders that are affected by others' expression.[66]

I think negative freedom of expression is philosophically well-grounded, but cannot resolve our questions of how to shape communication and distribute attention. Positive freedom of expression potentially offers more (though imperfect) practical guidance; at its most plausible, however, it is strictly speaking neither about freedom nor limited to expression. The concept of communicative justice does a better job of describing the normative concerns at which it gestures.[67]

---

[63] Ramsey, 1961.
[64] Lazar, 2023, 2024.
[65] Scanlon, 1978.
[66] Many thanks to Leif Wenar and Andrew Kenyon for helping me to see the force of the positive account. For important examples of philosophical engagement with online speech, see the essays collected in Brison and Gelber, 2019; Jongepier and Klenk, 2022.

Let's consider, first, theories of negative freedom of expression. In my view, the right to freedom of expression is most compelling when given deep foundations in liberal theories of individual autonomy or self-sovereignty. This view has fallen out of favour among specialists in this area, at least since Scanlon's own departure from his early work.[68] But I think it deserves a comeback—and that developing a theory of communicative justice can help revive it.

First, here's my preferred account. In general, the greater our de facto unilateral control over a sphere of action, the stronger the presumption that we *should* be entitled to exercise that control. I alone can control my body, my thoughts, my values, so there is a strong presumption in favour of me alone (or at least primarily) having control over each.[69] My de facto control over myself justifies de jure sovereignty over myself. Why is this? The elements of myself over which I have this kind of presumptive unilateral control are fundamentally part of my identity; to assert authority over them is to assert authority over what is core to me, and to force me to be a different kind of person than I would otherwise choose to be.

On these grounds, the core of the right to freedom of expression is a basic, fundamental right, as unimpeachable as the basic right to bodily integrity, or to freedom of thought. This right is, of course, not without limitations. I do not mean to wade into the vast literature limning the boundaries of this core principle of self-sovereignty. But, however those boundaries are established, there should remain a basic right to self-expression grounded in the value of individual autonomy. Any account of the political philosophy of online communication must take this right into account.

However, as important as it is, this right alone cannot provide guidance in shaping public communication and distributing attention. Theories of negative freedom of expression rightly focus on articulating its limits. They describe both the boundaries of permissible speech and the limits of permissible intervention in speech. Platforms could robustly secure negative freedom of expression while adopting one of a functionally infinite array of choices for how to enable people to express themselves, and whom to allow them to reach. Of course, they must also decide when to limit or penalise expression, but even then their ability to restrict people's expression is in fact rather limited. Individual platforms govern communication, not expression: they can determine whether your message reaches

---

[67] As an illustration of this point, consider the invaluable chapters of Brison and Gelber, 2019. As important as this groundbreaking book is, it offers little guidance on how to shape communication and distribute attention, beyond focusing on how the internet enables new kinds of speech-based harms that offer grounds to resource or curtail free expression. This is an important topic! But it does not exhaust the domain of communicative justice.

[68] Scanlon departed from his earlier (Scanlon, 1972) approach to freedom of expression, grounded more directly in autonomy, in favour of an interests based approach in Scanlon, 1978.

[69] Lazar, 2019.

*a particular* audience, not whether it reaches *some* audience, or whether you are able to express it.[70]

Theories of negative freedom of expression presuppose that powerful actors like the state can choose whether or not to intervene in expression, hence our task is to dictate when intervention is permissible and impermissible. In the terms of Lecture I, their approach is one of *extrinsic* governance — they aim to provide appropriate starting conditions for a healthy public sphere to emerge, and permit and enable unmediated communication. They seek to govern online communication the way a river's banks govern the flow of water. But algorithmic intermediaries must govern the digital public sphere the way chemical bonds govern the water molecules that they hold together. For online platforms, non-intervention is not an option. They unavoidably construct the medium of communication, which inescapably encourages some ways of communicating and undermines others. They unavoidably exercise intermediary power.

Suppose that a central authority had to decide what language a community would speak. Absent this decision, people would be incapable of communicating outside of their immediate family circles. And suppose that the language people speak could be adapted and adjusted in an incredibly fine-grained way, and that the central authority could monitor how the language was being used, and the degree to which it served its purpose, and then update it accordingly with relatively low transaction costs. An update once sent would lead to everyone in the community using the new version of the language. For this central authority the ideals of freedom of expression would be, if not entirely useless, at best a small part of the picture. The authority would share some responsibility for every communicative pathology that their new language enables, and every achievable communicative good that it fails to achieve. It needs to know not only what it must avoid, but also how it should positively be used.

Digital platforms are in basically this situation. They govern aspects of online communication that it would never be feasible to govern offline. Admonishments to not intervene are meaningless when every option involves intervening. Platforms cannot trust that a healthy public sphere will emerge from unmediated communication, because they cannot avoid mediating communication. They have to decide how to design the digital public sphere, not just wait for it to emerge.

In addition, the normative foundations of negative freedom of expression — the individual's self-sovereign right against interference by others — are also less salient in a highly constructed communicative context. For A to express herself, she needs (at that moment at least) little or no help from anybody else.[71] For A to communicate, she needs some other party to pay attention. And for A to communicate online, she needs other parties to pay attention, *and* some kind of digital intermediary to carry her message. Communication is necessarily social in a way that expression need not be. And online communication depends on intervening infrastructure, in ways that at least some offline communication need not. For A to communicate online, therefore, she must enlist the involvement of both her audience and the intermediary. It makes sense for theories of freedom of expression to focus on negative rights against interference by others, given the self-sovereign nature of self-expression. A theory of communicative justice, however, must focus at least as much on positive duties to appropriately shape others' communicative options and to distribute attention.

---

[70] This is, I think, the central idea in Renee DiResta's aphorism that 'freedom of speech' is not the same as 'freedom of reach'. See [DiResta, 2018](#).
[71] Obviously offline communication depends on language, which is a social product. However, language is not in any agent's control in the way that online platforms are.

Last, as is now widely recognised by freedom of expression scholars, in practice an excessive focus on negative freedom of expression likely shares some of the blame for the digital public sphere's pathologies. Research has long shown that, in online communication, protections for freedom of speech actually undermine many people's ability to express themselves, and actively thwart our attempts to build a healthy digital public sphere.[72] Digital platforms shape power relations between their users. Increasing users' freedom of expression means also increasing their ability to deceive, abuse and manipulate without consequences. This gives power to the deceivers, abusers, and manipulators. Worse still, arguments from freedom of expression are likely to be mobilised to defend platforms' right to manage online speech however they want, since their decisions about which kinds of communication to enable and how to distribute attention could be viewed as their own forms of self-expression, with which the state should not interfere.[73]

Many advocates of positive freedom of expression share these concerns about their negative counterparts.[74] They might think that we can mobilise the value of freedom of expression to better address the challenges posed by governance of the digital public sphere—for example by arguing that we should focus more on audience and bystander interests in expression than on the speaker's interests, and on the *positive* freedom to express oneself, which implicates a right to some kind of audience.[75] While I think this is headed in the right direction, I think that 'freedom of expression' is the wrong normative lens to adopt: the answers we seek lie in neither expression, nor in freedom, alone.

If you think that speakers have basic rights not just to be free to express themselves, but to reach an audience (some particular audience, or an audience in general), then you are arguing for a *communicative* right, not a right of expression. Indeed, you are probably arguing for a suite of distinct communicative rights, each indexed to a particular audience. If you aim to ensure that speakers do not pollute an audience's information environment, directing their attention to misinformation, then your primary concern is again with wrongful communication, not with wrongful expression. If a liar pipes misinformation into the ether, and no audience attends to it, then no harm has been done.

And *freedom* is not the only value that matters. Suppose we had to choose between two competing arrangements for the digital public sphere, each of which limits (and supports) individuals' *freedom* of expression to the same degree. If freedom of expression were all that matters, then we should be indifferent between these two

---

arrangements. But what if, within those identical limitations, they encourage and discourage different practices of communication? And what if two arrangements give speakers access to quite different audiences, and result in quite different distributions of attention? Focusing on freedom of expression treats the challenge here as one of extrinsic governance—set baseline parameters within which people are then free to operate in an unmediated way. But as argued in Lecture I, algorithmic intermediaries also perform *intermediary* governance, they constitute the social relations that they mediate. This means we have to attend to much more than just settling baseline or background parameters—even when that baseline is enriched by a positive theory of freedom, rather than a thinner negative one.

Consider the case of the public language above. However important the value of positive freedom of expression, it does not, without conceptual jerrymandering, cover the full gamut of what is entailed by deciding on a modality of public communication and a distribution of attention. Besides freedom, we also care about promoting the other interests that communication serves; we care about equality; we care about being collectively self-governing. Freedom is one value among others, with which it potentially competes.

No doubt there are sometimes strategic, pragmatic reasons for stretching the concept of freedom of expression so that it can cover this broader domain of normative concern—perhaps as an exercise in re-interpreting the US constitution, to put its first amendment to less libertarian use. However, as a matter of political philosophy we can set these pragmatic considerations aside, and call a spade a spade. If we want principles to guide how platforms shape communication and distribute attention, then freedom of expression must only be part of the picture. In fact, if we can offer a broader account of the morality of communication, then we can I think resuscitate the core, autonomy-based right to freedom of expression that has fallen out of favour. Critics of the autonomy-based account have (I suggest) made a subtle category error. They have rejected a well-grounded principle because it was not capacious enough for their normative ambitions. But the problem is not with the well-grounded principle, but with the attempt to use freedom of expression to cater for all of those ambitions. We need an account of the morality of communication more broadly, and we must attend to other values besides freedom: I think we need a theory of communicative *justice*. And an autonomy-based right to freedom of expression can take its place as one element in such a theory.

However, why call it communicative *justice*? Why not communicative freedom, or communicative equality or some other such good? For three reasons.

First, minimally, when expression is so easy as to be near costless, attention is the scarce resource.[76] The distribution of a scarce resource among moral equals is a problem of justice.

Second, and more generally, demands of justice arise when moral equals seek to realise individual and collective goods through cooperative action that requires restraint and coordination, and therefore yields benefits and burdens. Principles of justice determine how we can act so as to realise those goods in ways that appropriately reflect our fundamental moral equality, in part by ensuring that the benefits and burdens of realising that 'cooperative surplus' are fairly distributed.

---

[76] Wu, 2017; Pedersen *et al.*, 2021.

Our practices of public communication take just this form. They necessarily require coordination and collective action (you cannot communicate alone). They enable us to realise individual and collective goods that we would not otherwise attain. Successful public communication requires us to sometimes show restraint, and it involves benefits and burdens that can be distributed in better or worse ways. The principles shaping how we can pursue the fulfilment of individual and collective communicative goods in ways that appropriately respect our fundamental moral equality, by among other things ensuring that the benefits and burdens of realising those goods are fairly distributed, are principles of justice.

Third, at least in *A Theory of Justice*, Rawls construed justice less as a discrete normative concept, and more as a way of finding an optimal balance between other more fundamental values.[77] In particular, Rawls' two principles of justice articulate commitments to freedom, fair equality of opportunity, distributive equality, and the promotion of individual well-being (as measured in primary goods).[78] I understand communicative justice in a similar way: a theory of communicative justice should say how to promote people's individual and collective communicative interests in ways that respect their fundamental status as moral equals, by articulating values like individual freedom, relational equality, and collective self-determination. Justice is the right word for this complex articulation of complementary commitments.

But might one not complain about the risk of 'justice inflation', where the peremptory urgency of justice is too liberally invoked to address lesser normative concerns? We need to distinguish, here, between two ways in which 'justice' can be invoked. Justice is sometimes used to describe a set of minimum standards, criteria beneath which society must not be allowed to fall. And sometimes, as in Rawls, it is invoked to describe the 'first virtue of social institutions'—an ideal that they may never attain. I am using the idea of communicative justice in this second, aspirational sense.

## 4.   THE CURRENCY OF COMMUNICATIVE JUSTICE

A theory of justice should comprise at least two things: an account of the goods at stake (whether as objects of pursuit, or as a scarce resource to be distributed), and an account of the norms by which the pursuit of that good must be constrained, to ensure that benefits and burdens are fairly shared. In this section, I will describe the *currency* of communicative justice. In the following section, I will argue for a set of norms of communicative justice. I will be partisan: I will defend a theory of communicative justice that unifies underlying commitments to liberty, relational equality, and collective self-determination that I introduced in Lecture I. But my aim is less to convince you of this particular theory of communicative justice, more to open up this terrain for further inquiry by political philosophers.

I'll call the goods promoted by communication *communicative interests*. And I will entertain three categories: non-instrumental individual interests; instrumental individual interests; and collective interests. In practice these overlap quite well with T. M. Scanlon's noted division between participant, audience, and bystander/third-party interests in expression.[79] However, Scanlon divided things in this way because his core task was to determine when an individual's interest in speaking X should be overridden by his audience or some third parties' interests in his not speaking X. My task is instead to think about the individual and collective goods that our communicative practices serve. So it makes more sense to divide them up based on the kinds of interests they are, rather than based on whom they belong to, especially since the digital public sphere now enables mass multipolar communication, so we all rotate through speaker, audience, and third party roles, or else occupy them at the same time.

---

[77] In later work Rawls argued that justice is equivalent to fairness, but even then fairness remained an articulation of these other values. See Rawls and Kelly (Ed.), 2001.

[78] Rawls' defence of the 'difference principle' is, as Cohen argues, fundamentally a way of balancing equality with the promotion of overall well-being. See Cohen, 2000.

Communication is a compound of expression and attention. A communicates with B when A expresses himself, and B pays attention. The basic expressive interest is a desire to make one's ideas manifest in the world. No doubt this is sometimes *purely* expressive—an artist compelled to create may be indifferent whether another person ever views their work.[80] But for most of us, what matters is not simply catharsis but connection: we express ourselves because we hope an audience will understand and appreciate what we say, and recognise in us the potential that our self-expression imperfectly captures. To be listened to, to be seen: we value these things partly because receiving an audience amounts to an acknowledgement, however minimal, of our shared membership in the moral community.[81] And to be ignored is to feel like an outcast, no member of the moral community at all.

To be listened to and seen is to be acknowledged as a person, a source of ideas, opinions, and creativity. To have others devote genuine effort to attending to your self-expression, to put in the time to follow and understand your work, is an ineffable privilege.[82] Mere acknowledgement is sufficient for the fundamental respect that is our due as members of the same moral community. Sustained and costly attention is a mark of the esteem (which many though not all of us seek) that goes beyond that moral minimum.

Mere expression needs no audience. One-way communication derives its value from the significance of being acknowledged as someone who cannot be ignored, or esteemed as someone worth attending to at length. But surely the most significant non-instrumental communicative interest is in true two-way communication, in which both sides express themselves, attend to the other, and respond. Seeing and being seen, listening and being listened to: this joint activity makes one's life go better just through its exercise. To participate in a healthy conversation is to create something together. This is conditionally non-instrumentally valuable: if your end is evil, then that you acted together towards it is no saving grace. But the value of acting and conversing together is not reducible to the good thereby realised. Acting together to a trivial or silly end (such as a whimsical conversation) is non-instrumentally valuable just because of the cooperation involved. This mutually-recognising communication most often happens in small groups, but can also involve larger groups, even a whole society

---

[79] Scanlon, 1978
[80] 'Expression often has nothing to do with communication' (Cohen, 1993: 224).
[81] Honneth and Fraser, 2003.
[82] One of which I am eminently aware in writing this very long essay.

under favourable conditions.

We also have an interest in others communicating with us as equals. In modern societies, our physical or material interactions are relatively limited—equality is primarily served by non-interference, and by contributing to collective projects that realise more ambitious goals, e.g. by paying taxes. But our scope for interacting with one another through communication is much greater—simply put, talk is cheap; and we are now able to communicate with anyone, anywhere in the world, trivially cheaply. Indeed, our social relations with one another are fundamentally constituted, at least in part, by our communicative practices. So our practices of communication afford one of the principal positive ways we can affirm our fundamental relational equality.

Indeed, some of our most essential acts delineating the contours of our moral obligations to others—consent and contract—are fundamentally communicative.[83] The intrinsically ethical nature of communication has led some to seek foundations for morality itself in a set of idealised communicative practices, or in the notion of conversibility that successful communication entails.[84] But we needn't rise to such heights to see the key point here. If we are interested in relating to one another as equals, and if our social relations are in part constituted by communication, then we have an interest in communicating with one another as equals. In practice this means at least going beyond minimally acknowledging someone as a speaker, though perhaps falling short of the deep engagement constitutive of high esteem. It means taking people seriously as speakers.

Certain kinds of communication are constitutive of a life going well: being acknowledged; being attended to; participating in the give-and-take of true bilateral or multilateral conversation; communicating with others as an equal. But communication is also instrumental to almost all the other goods life offers. I want to highlight four broad categories: knowledge, coordination and collective action, individual and collective identity formation, and entertainment.

Communication allows us not only to consume information, but also to think together with others, to be challenged by them, and to be surprised. Communication is arguably essential for acquiring moral insight (or knowledge) and is undoubtedly indispensable for reaching a better understanding of the societies of which we are part. Successful governance of communication can create a healthy information environment; failed governance leads to epistemic pollution. That this point is obvious, and can be quickly stated, should not undersell its centrality to our communicative interests.

And, of course, communication is the sine qua non for resolving the coordination and collective action problems that plague any attempt to live together in society with others. Communication is our means to signal our willingness to cooperate with others, to deliberate and form plans, to bind ourselves through promises and contracts, and to license others to act in ways only permissible through consent. Our lives have always depended on our ability to act successfully with others,

---

which depends on our communicative practices.[85]

Communication shapes individual and collective identity: one's sense of the kind of person one is and the community to which one belongs.[86] This communicative interest has been served especially well by the transition in attention from mass to social media, as the departure from programming for the median viewer in favour of meeting subcommunities where they are and, in general, the 'wide aperture' of social media have enabled minoritized social groups to discover one another and come to a shared understanding of their identity.[87] This ability to reach a sense of one's place in the world is invaluable for anyone, but is especially important for those whose other opportunities to find their community are otherwise limited.[88]

Amid the more worthy aspirations of liberal political philosophy, it is easy to forget that one central role of communication in our lives is simply to give us pleasure—to entertain us. One might be tempted to hive off entertainment from other considerations that contribute to communicative justice as being the wrong kind of good to pursue in this moralised way. But entertainment itself raises deep questions of justice (for example, as concerns which social positions are represented and how). And entertainment and political communication are always in close dialogue with one another, and often directly overlap. We have a communicative interest in being part of a vibrant creative economy.

Collective goods have some or all of the following features: individuals enjoy them in virtue of their membership in some relevant collective; they are public goods, provision of which by some members of a collective secures them for other members, whether or not they seek them out; they are irreducibly social goods, where one's having the good is conditional on other members of the collective also enjoying it.[89] National self-determination is the paradigmatic example: I enjoy it in virtue of being Australian; every Australian enjoys it whether or not they have worked to realise or defend it; and whether I benefit from it depends at least in part on whether a sufficient number of my co-citizens benefit from it too. The most crucial collective good at stake in the pursuit of communicative justice is what I will call *civic robustness*.

I owe this idea to the tradition of work on the public sphere, which (for me) begins with John Dewey, and passes through Habermas and Iris Marion Young to, most recently, work by Joshua Cohen and Archon Fung on the digital public sphere.[90] Dewey argued that the putatively private decisions of individuals inevitably cause negative externalities for others; the central political challenge is for those affected by these externalities to unite as a public and set boundaries within which private

---

exchange may proceed without unduly harmful effects on others.[91]

We need to expand on Dewey's idea in three ways. First, the animating impulse behind a public's formation is not simply the existence of harmful externalities. Instead, it is the exercise of, especially, *governing power* by some over others. These power relations call for the public to emerge as one of the principal means by which power can be directed, limited and corrected. The presence of a vocal public that draws attention to the decisions of the powerful and criticises them when needed is vital to accountability.

Second, because power relations call for publics, we should not think only of *the* public or consider the nation-state the only relevant public.[92] Of course, the power of the nation-state is unequalled, so if we need a robust public anywhere, we need it there. But civic publics, as I will call them, are needed wherever significant power — especially governing power — is exercised.[93] Think, for example, of transnational organisations like large technology corporations, or subnational organisations like universities and other employers. A healthy digital public sphere enables civic publics to emerge at these supra- and subnational levels too. Indeed, while we rightly lament the shortcomings of the digital public sphere, it has proved *very* effective at generating civic publics responding to the transnational power of big tech companies, resulting in the 'techlash' and materially contributing to significant new regulations in the EU, and most likely in other major economies too.

Third, as I. M. Young argues, while civic publics must hold the powerful to account, they should also be sites of creativity and innovation, where new ideas are forged that the public can pick up and, if not implement directly, at least use to create the foundations for implementation.[94]

To bring these threads together in an attempted definition of civic robustness: a public sphere is civically robust if and to the extent that it supports the emergence of effective civic publics in response to the exercise of governing power. Civic robustness is a collective good because a civically robust public sphere limits power (especially governing power) and provides the discursive foundations for successful collective action. These are goods we enjoy, as members of a civically robust society, in virtue of that membership.

Other more formal constraints on power matter too. But civic robustness need not rely on implementation by *other* powerful agents to have effect. It cannot be straightforwardly co-opted or corrupted. It can limit power even when other means fail. This protects our basic liberties and serves relational equality. If you are subject to power, but that power is held to account by a civic public in which you

---

[91] 'Thus perception generates a common interest; that is, those affected by the consequences are perforce concerned in conduct of all those who along with themselves share in bringing about the results' (Dewey, 2016 (1926): 84).

[92] Fraser, 1990 introduced the idea of subaltern publics, updated by Squires, 2002. Although subaltern publics are clearly important, I mean the concept of civic public to be defined not by the ascriptive characteristics of its constituents, but by the power structures in response to which it emerges.

[93] Young, 2000: 178-9.

[94] 'The public sphere is the primary connector between people and power. We should judge the health of a public sphere by how well it functions as a space of opposition and accountability, on the one hand, and policy influence, on the other' (Young, 2000: 173).

participate as an equal, then you have (some) power over those who have power over you.

Civic robustness is not sufficient for collective action. A civic public cannot simply translate public opinion into practical outcomes. That requires material power. But the exercise of material power to achieve positive goals relies partly on inspiration from the public to set a course of action. And it depends on public endorsement, not only of the agency with material power, but also of the others with whom we are acting in concert. Collective action requires mutual commitment, grounded in a public expression of the willingness to act together (and non-expression of the refusal to do so). These are the *discursive foundations of collective action;* they are necessary for us to act together to realise social and structural change. This serves the value of collective self-determination and bridges significant gaps in its realisation by the institutions of representative democracy.[95]

Civic robustness is generally considered valuable because of how it contributes to successful representative or deliberative democracies. While civic robustness clearly is an important democratic value, I think it should not be assessed solely through the lens of democratic theory.[96] It is, instead, a *detachable* complement to democratic institutions.

The underlying idea of civic robustness is fundamentally democratic since it concerns the subjects of power uniting to shape how that power is used. But a state could be a healthy democracy and yet lack civic robustness, if it is well run, suitably constrained by robust procedures, and disagreement over the state's direction is relatively mild. And a monarchy or other form of 'benevolent dictatorship' could in principle enjoy a strong digital public sphere, where the unrepresentative government is held to account, and ideas are floated for adoption by the government. And civic publics are needed wherever power is exercised, even in contexts where democratic governance is either not feasible or not desirable—for example in response to transnational or subnational power. My university is not, in any respect, democratic—and perhaps it should not be. But it clearly *is* exercising governing power over students and employees, and civic robustness is a vital counterweight to that power.

Civic robustness is democratic in giving some power to the people, but it does not entail the people *ruling*. Holding to account is not ruling, nor is providing the discursive foundations for collective action. To *take* collective action, you must have the levers of material power. That requires more than civic robustness; it requires full democracy.

This observation has important upshots. First, existing philosophical work on the digital public sphere may err in assimilating it too closely to theories of deliberative democracy. Civic robustness is a distinct ideal from the discursive goals of deliberative democrats.

Second, assessments of the digital public sphere should not collapse into evaluations of the health of democracy. Democracy is in peril for many other reasons besides the pathologies of the digital public sphere. We can determine civic robustness independently from our assessment of democracy. And while a civically robust digital public sphere is invaluable in part because it contributes to a healthy and vibrant democracy, that is not the only reason why it matters.[97]

---

[95] Young, 2000: Chapter 4
[96] This tendency is widespread. See e.g. Persily and Tucker, 2020; Fukuyama, 2021; Haidt, 2022.

Civic robustness is both instrumentally and non-instrumentally valuable. It expresses our basic equality and capacity for collective self-determination; it enables us to protect our basic liberties and is instrumental to achieving other goods. Beyond that, civic robustness is a public good—because non-excludable and non-rivalrous, as well as a primary good—because whatever else you want as a society, civic robustness is likely to help you get it. It is also an irreducibly social good: you cannot enjoy civic robustness in isolation from the other members of the salient civic public—the ability to lay the discursive foundations of collective action relies on there being a collective that can act together. This implies that a civic public should be inclusive of all those with a claim to participate. If you have created a sectarian civic public which influences the exercise of power to benefit your sectarian group, while it remains unaccountable with respect to others, then you haven't contributed to civic robustness, you have simply co-opted the levers of power to your group's advantage.

## 5.   NORMS OF COMMUNICATIVE JUSTICE

Algorithmic intermediaries that govern the digital public sphere should shape public communication and distribute attention so that it advances these individual and collective communicative interests. A consequentialist approach to public communication would perhaps stop there, and argue that the sole target should be to maximise the fulfilment of those communicative interests. But this would be deeply at odds with the fact that the members of the community whose communication is being shaped and whose attention is being distributed are moral equals, who have claims to certain kinds of fair treatment even when it is not optimific. What does it mean to promote these communicative interests in ways that respect our underlying moral equality?

I argued in Lecture I that the exercise of governing power has to be justified against three distinct standards: substantive justification, proper authority, and procedural legitimacy. The first, 'what', question concerns the ends that power is being used to serve. The second, 'who', question, concerns whether governing power is being exercised by those with the right to do so. The third, 'how', question concerns whether the manner in which power is being exercised is defensible. Realising communicative justice means justifying the promotion of communicative interests against each of these three standards.

---

[97] Recognising the difference between the value of civic robustness and of democracy also undermines observations like this from Benkler et al.: 'asking platforms to solve the fundamental political and institutional breakdown represented by the asymmetric polarisation of the American polity is neither feasible nor normatively attractive.' That seems right: blaming platforms for the collapse of American democracy is excessive, and holding them accountable for fixing it seems a mistake too. But online platforms *have* undermined civic robustness in the digital public sphere; since they essentially constitute the digital public sphere, they have a responsibility to address that. Benkler *et al.*, 2018: 367.

## 5.1. What: Reasonable Disagreement

Describing communicative interests at a high level, from a particular perspective, is one thing. Algorithmic intermediaries that shape communication and distribute attention have to optimise for some particular set of communicative interests, in circumstances of radical disagreement. The first step in advancing our communicative interests in an egalitarian way is to step back and ensure that our understanding of precisely what we are aiming at is sufficiently cognisant of the range of reasonable disagreement about why communication matters.

One approach would be for platforms to simply decide what they think is in everyone's communicative interests, and then promote that—this would obviously be objectionable, as it would amount to a private company deciding what counts as acceptable or attention-worthy communication. Another approach would be to punt to users' preferences, and argue that people are the best judges of what's in their own communicative interests, so we should simply give them what they want. But this is precisely the tried-tested-and-failed approach of optimising for engagement, which has led us to the very pathologies that we are seeking to transcend. Moreover, placing so much weight on users' preferences seems seriously unwise, given that their preferences are often endogenous to the platforms they use.[98]

An alternative response would be for platforms to abandon the goal of trying to directly promote our communicative interests, and instead find ways of shaping communication and distributing attention that are justificatorily neutral, that don't involve making any substantive judgments about what kinds of communication are beneficial for us. As argued in Lecture I this is obviously hard, if not impossible, to do; but some might argue that they should at least try, and that they have at least one lever on which they can pull: they can abandon the paradigm of actively *distributing* content, and instead either adopt a reverse chronological feed, or else simply enable direct communication (as on Discord, for example), without any feed.

Reverse chronological ordering of online content has one clear moral advantage over the algorithmically-curated feed: it is *content-independent*, therefore it does not involve making any judgments as to the relative merits of different posts. This does matter: when some place themselves in a position to make judgments of what is good or bad for others, they thereby place themselves above those others, undermining relations of equality between them. In this respect, reverse chronological feeds are more egalitarian than algorithmically curated ones. They get closer to justificatory neutrality than other approaches.

However, this means of preserving equality in one respect comes at a significant cost to our communicative interests. Reverse chronological feeds abdicate responsibility for curation, and ultimately prove tractable only insofar as the costs of filtering and ranking what one sees online are devolved to the user. This is not only incurably tedious (though perhaps this is just me), it also means that we are ultimately likely to reproduce the pathologies of applying the 'consumer mindset' to the distribution of online attention.[99] It radically limits our ability to curb the distribution of communication that undermines our communicative interests, and,

---

critically, cripples the ability of online platforms to direct collective attention in ways that help publics find themselves. Reverse chronological ordering *fetishizes* neutrality, prioritising it above all else, and foregoing the opportunity to aim for other dimensions of communicative justice out of excessive and ultimately inconsistent opposition to taking a moral stand. It might offer a content-independent approach to filtering and ranking, but platforms must still make value judgments concerning how to shape communication in every other respect of platform design and moderation.

The commitment to neutrality is ultimately grounded (as I argued in Lecture I) in values of liberty, equality, and collective self-determination. We care about neutrality because: we don't want people limiting our options based on their own moral views, whatever they might be; because imposing your values on others implies lack of respect for their equal capacity to reach their own reasonable moral views; and because if we must make decisions about how to live together (as we must) we should take those decisions together. Using recommender systems to shape public communication and distribute attention in line with the ideals of communicative justice advances these goals. If we fetishize neutrality, we are sure to not only fall short of these ideals, but to facilitate communicative injustice. In particular, the capacity of algorithmic intermediaries to allocate collective attention and so help civic publics to find themselves could be an invaluable counterweight to unaccountable power. We should give up this power only if we are convinced it cannot be used appropriately. We should therefore try to find other ways to address the problem of reasonable disagreement, rather than tying one hand behind our backs in the attempt to realise communicative justice. We should aim for liberal pluralism rather than aspiring to an unattainable and undesirable justificatory neutrality.

The problem of reasonable disagreement is in part a problem of substantive justification, and in part one of authority. I address the authority part below. The best remedy for the substantive problem is not to abdicate responsibility for judging what will advance people's communicative interests, but to advocate for a conception of our communicative interests that is as pluralist as possible, given background disagreement. In particular, we should focus attention on communicative interests that are plausibly primary goods, in the Rawlsian sense that you benefit from them whatever your other, more fundamental goals are in life.[100] The collective communicative good of civic robustness is clearly a primary good, as argued above. The same is true for our instrumental communicative interests in knowledge formation, coordination and collective action, and the opportunity to forge one's identity among one's cognates. The interest in a vibrant creative economy is perhaps more borderline.

Our non-instrumental communicative interests are less obviously primary goods. But being acknowledged and esteemed, participating in the fruitful joint action of a healthy conversation, communicating with others as equals and so avoiding deception, abuse and manipulation, are all fundamental human interests, as well as plausibly being constituent elements of what Rawls called 'the social bases of self-respect'.[101] Importantly, they are not tied to any particular broader conception of the good life, and in particular seem independent of one's political leanings.

---

[100] Rawls, 1999: Section 15.
[101] Rawls, 1999: Section 67

Whether you vote left or right, for example, these communicative goods have value.

Of course, determining just which communicative practices fulfil these interests—for example, which count as deceptive or manipulative—is inevitably going to be contentious, and liberal pluralism at this level is likely to be hard to achieve. Famously, despite his best efforts Rawls has been criticised for explicitly favouring some ways of life over others.[102] Any aspiration to neutrality always leaves a justificatory deficit, which must be filled with democratic authorisation—my answer to the 'who' question. But we should still *aim* for pluralism, so that the goods we are striving to achieve can be recognised as worthwhile by all those with a stake. It is still other things equal better to be (reasonably) pluralist, and advocate for goods that most people can get on board with, than simply to ride roughshod over their disagreements.

Digital platforms should shape communication and distribute attention in ways that incentivise and encourage the fulfilment of our communicative interests, and frustrate and discourage communication that thwarts them. They should operationalise these interests in a measurable way, monitor their platforms to track the degree to which they are achieving this goal, and adapt to improve performance. Each of recognition, esteem, dialogue, equal treatment, knowledge, deliberation, coordination, community and entertainment can be actively designed for, or designed against. Algorithmic intermediaries at present do a poor job of fulfilling these interests. The design of communication options, safety and moderation practices, and the distribution of attention should be oriented around doing better. Understanding precisely which interventions will serve these ends and which will frustrate them is undoubtedly very challenging in practice.[103] But the malleability of digital communications technologies and the ability to experiment with real-time feedback can prove invaluable.

Operationalising each of these instrumental and non-instrumental individual communicative interests is undoubtedly challenging, but they can at least be promoted directly. Realising civic robustness requires a few more intermediate steps.

First, civic publics cannot perform either of their primary functions—holding power to account and providing the discursive foundations for collective action—if they lack access to accurate, relevant information about the powerful agents they are holding to account' about the consequences of their actions, and about the need and opportunities for collective action. This means realising a healthy information environment, as Sunstein and others argued,[104] and as is already necessary in order to fulfil the individual communicative interest in coming to better understand the world. Notice, though, that thinking about civic robustness forces us to recognise that a healthy information environment is not only an individual good; it is also a collective one. Your information environment contributes to civic robustness only if it is healthy not just for some, but for all (or nearly all). If you can access good information, but others in your community are too easily led astray, your information environment is unhealthy, and civic robustness is at risk. Indeed, if

---

you have good information, but it even *appears* that other members of the relevant civic public are systematically deceived, then realising the kind of trust and mutual toleration necessary for civic robustness may be unattainable.[105] Civic robustness relies not only on our having access to good information, but on it being common knowledge both that we do and that we use that access to form broadly reasonable beliefs.

This will ground decisions about platform architecture and recommendations—for example, we should (obviously) design online platforms to reward the production of high informational value news content rather than incentivising clickbait, sensationalism, and extremism.[106] But it also implies that norms of communicative justice go beyond the platform—we, as societies, and the companies profiting from platforms, likely have special obligations to remedy the harm to the news media industry done by the advent of social media.[107]

Second, as Habermas argues the success of the public sphere depends on the availability of a well-protected *private* sphere wherein people can communicate securely as they form the ideas and coalitions that ultimately play such a significant role in the public sphere.[108] Realising communicative justice in the digital public sphere likewise also entails securing sufficient scope for private digital communication. The availability of private means of end-to-end encrypted communication might seem orthogonal to the values of communicative justice that focus on *public* communication. But we likely cannot enjoy civic robustness without these robust protections for private communication.

Third, for civic publics to come together in response to the exercise of power and to lay the discursive foundations for collective action, they must be able to find themselves. In *The Public and its Problems*, Dewey argues that the industrial revolution radically increased externalities caused by private actors' profit-seeking activity, without seeming to create an equivalent increase in the ability of those affected by those externalities to come together and act in response. He argued that we need a 'Great Community' to hold the 'Great Society' to account, but that the central challenge for that community to take form, was for people to find one another and direct their attention to a common purpose.[109] The allocation of collective attention—where many eyes are on the same thing, and this fact is common knowledge, in part because people are engaging *en masse* with the object of their attention—is crucial to forming civic publics and giving them sway over those they hold to account. Digital platforms play a decisive role in allocating collective attention. This sometimes has negative results, for example when some unfortunate becomes social media's 'Main Character' for the day. And even its positive effects are fragile and insufficient—as Zeynep Tufecki argues, activism and political movements born on (then) Twitter are often ineffective in the end.[110] Nonetheless, the distribution of collective attention is essential for a civic public to find itself—it's how we know who else is concerned and might act together with us. And the allocation of collective attention is then essential for the civic public to achieve its goals, holding the powerful to account and laying discursive foundations for collective action. Even if we have not yet determined how to use algorithmic and other means of curation to optimally allocate collective attention, even more than in Dewey's day we now have 'the physical means of communication as never before', and online platforms have the *potential* to truly enable civic publics to find themselves. Understanding how to unlock that potential is among the most urgent challenges of our time.

Fourth, and most important, for civic publics to perform their functions, the public must not only find itself but must hold together. This relies on participants having and displaying certain attitudes towards each other. They must tolerate one another well enough to be able to deliberate on how power is being exercised and what course of action to take.[111] And in the inevitable event that they do not unanimously agree on the right course of action, they must be willing and able to both commit themselves to a joint course pending some appropriate decision procedure followed by those with material (not just communicative) power, and to trust one another to abide by those commitments.[112] Communicative justice norms must therefore promote mutual toleration and trust and reduce their contraries. This will be a subtle art—toleration and trust might best be promoted indirectly. But we clearly can and do communicate with others in ways that undermine mutual trust and toleration, and we can do better. Platforms must shape public communication and distribute attention to encourage the latter.

Note the contrast between this approach to communication in the digital public sphere and an alternative that focuses not just on the necessity of trust and toleration but more narrowly on how people engage one another in public debate.[113] Cohen and Fung argue that we should encourage a norm (not an outright requirement) whereby people express themselves in the digital public sphere in a certain way. People should justify their arguments to others in terms of a common good, rather than just hoping to win enough support for their sectional interest that they can proceed over opponents' objections.[114]

These kinds of discursive norms—even in the attenuated form advocated by Cohen and Fung—have a chequered history. The ability to present one's case in public as a reasoned argument grounded in the common good may not be a

---

straightforward function of the reasonableness of one's position or one's broader civic virtue. Rhetorical invention and superficial cleverness might be just as prominent causal explanations, and we can generally assume that those already privileged will find it easier to meet these constraints than those who seek to overturn that privilege.[115]

More generally, while these discursive norms might be facially appealing (even if exclusionary in practice), they are, on the whole, surplus to requirements. When theorising the digital public sphere, we should probably abandon higher ideals of public reason and deliberative decorum. We want people to tolerate one another well enough to talk and listen to each other, and to trust one another enough to be willing to commit in advance to supporting decision procedures whose outcome is uncertain. This is a kind of minimally egalitarian communication which recognises that the other is someone to be reckoned with, someone whose input counts. To promote it, we should actively promote toleration and mutual trust and weed out affordances of platform architecture and recommender systems that tell in the opposite direction (which is currently most of them).[116]

## 5.2. What: Baselines

If algorithmic intermediaries are to shape communication and distribute attention in ways that advance communicative interests, they must also sometimes use adverse interventions that prevent, punish, or at least disincentivise practices that undermine the fulfilment of communicative interests—these are the safety and moderation practices introduced above. A theory of communicative justice should account for when it is appropriate for algorithmic intermediaries to exercise these enforcement powers.

With some caveats, for this specific problem I think existing theories of permissible constraints on speech should provide adequate foundations. Of course, online platforms are not states. Indeed, states are often the biggest *threats* to communicative justice, and are powerfully incentivised to exceed any reasonable constraints on their interference into speech, and to force platforms to do the same. As a result, online platforms can sometimes be unlikely bulwarks for individual freedom against state power.

More importantly for present purposes, however, the stakes of content moderation are not the same as the stakes of state governance of speech. This is true both quantitatively and qualitatively. Quantitatively: states can penalise speech by depriving the speaker of property or liberty. Platforms can deprive you of an audience, and of the ability to participate in relevant civic publics. The stakes are lower. But the stakes are also qualitatively different. When states limit your self-expression they potentially violate a negative right—a right against interference by others. When platforms limit your ability to communicate with a particular audience on that platform, they potentially violate a positive right—a right that the

---

platform help you to achieve a particular end. Other things equal, negative rights enjoy more robust moral protections than do positive rights.

This helps answer one preliminary normative question: do platforms even have the right to remove communications that breach their policies?[117] The answer there seems pretty clearly yes. The contrary would entail that they have a positive duty to assist others in expressing themselves, no matter how obnoxiously they do so. One might indeed conclude that platforms should enjoy considerable latitude in deciding whether to help you reach an audience or not. Your presumptive claim to aid is grounded in weaker interests than your claim that others not interfere with your self-expression. And one might question whether they have *any* positive duties to help you reach an audience. Indeed, one could even argue, as alluded to above, that decisions about how to shape communication and distribute attention on a platform are themselves forms of self-expression by the platforms, so they cannot be required to adopt any particular policy. But one cannot plausibly argue that platforms lack the right to enforce well-grounded policies for appropriate communication *on their platform*.

The question of what platforms can be required to do, and by whom, is ultimately a question of authority. I return to it below. For present purposes, however, I note only that if you have built up an audience on a platform through good faith compliant behaviour over time, then you have a legitimate interest in continuing to be able to access that audience, which ought not be capriciously taken away. And we have communicative rights to participate on platforms that play an important role in civic publics that we have an (independent) claim to be part of. Of course we can, through our behaviour, forfeit the protection of these communicative rights, at least for a period. But even if they are positive rights, they still have significant weight.

As such, platforms should not take decisions to suspend or ban accounts lightly, and one should be able to derive a suitably stakes-adjusted account of the boundaries of permissible online speech from a broader theory of freedom of expression. Very roughly, this kind of enforcement should be applied only in response to, and in order to prevent, sufficiently serious communicative harms (where a communicative harm is a setback to a communicative interest).[118]

Matters become more interesting when we are forced to consider patterns of behaviour that our existing theories of the limits of free speech are ill-placed to address. For example, consider the categories of stochastic and collective harm.

To see the difference between them, consider these cases.[119] In the first, A is trapped in a tank that will be filled with water, drowning him, if a switch is flipped. A thousand people, the Bs, each have a button before them, and each button has a 1/1000 probability of causing that switch to flip, as well as some prospect of reward for B. Assume the probabilities are independent. If the Bs all press their buttons, and A is harmed, then that is a stochastic harm. In almost all cases, the Bs pressing their buttons was causally ineffective. For at least one B, it was not. But every button pushed risked harming A.

---

Now consider a second case, in which A will be drowned if the volume of water in the tank exceeds 700 litres. Suppose that each of the 1,000 Bs has a button before them that will, when pressed, pour one litre of water into A's tank, as well as give them some reward. If the Bs all press their buttons, then the resulting harm is a product of all of their actions (or at least, that of a subset of at least 700 strong). This is a collective harm, where each of the Bs contributed to the harm, none of them individually made a difference counterfactually speaking, but the harm could only be fully realised if a high enough number of them did so.

Online communication is rife with both stochastic and collective harms. For example, stochastic manipulation and stochastic radicalisation: many statements are made, each of which has a low probability of effectively manipulating or radicalising any given individual; but over a large enough population the probability that *someone* will be manipulated or radicalised gets very high.[120] Or consider collective harms caused by platform design and algorithmic amplification, as when individually innocuous 'likes' cumulatively imply widespread disrespect for the target of the 'liked' statement.[121]

Suppose, then, that a speaker, S1, performs a speech act X, and X raises the probability that someone—one of the As—will suffer some harm, and one of the As, A1, does in fact suffer harm, in part due to X and speech acts like it. Then S1 has contributed to a stochastic harm to A1. When S2's speech act Y contributes a small amount to a harm to A2 that many others also contribute to, such that these disparate, individually insignificant contributions amount to something serious when experienced together by A2, S2 has contributed to a collective harm to A2.

In these cases, by stipulation, S1 and S2's actions do not, in their own right, constitute sufficiently serious transgressions to justify penalising S1 and S2 according to established freedom of expression norms. X and Y are harmful, in the end, because of the pattern of similar behaviour by others that they contribute to. Indeed, X and Y might be either meaningless or utterly harmless in their own right, and derive their significance entirely from those patterns of behaviour and the broader intentions of others. And the platforms enable those patterns to emerge—they provide the context that knits together these different actions into a pattern. In the analogy above, the platform is equivalent to the water tank, in virtue of which the many different individual contributions realise a harmful outcome.

In such cases, extreme measures like suspending or banning S1 or S2 are not proportionate to their degree of guilt, or their degree of causal contribution. Cases like these are well-suited to filtering-as-moderation. Platforms should adopt policies that describe collective and stochastic harm, and should use filtering to reduce the degree to which X and Y contribute to and derive significance from the broader set of similar posts online. And they should do so not because of the liability of individual contributors to those stochastic and collective harms, but because they, the platform, are responsible for knitting together these individual actions into the harms in which they result. I return to the interesting question of whether such filtering should be done transparently below, in response to the 'how' question. As a first pass, given that the proper target of such interventions is the platform itself, this seems a case for platform observability rather than individual rights of due process.

## 5.3. What: Distribution

Realising communicative justice involves promoting communicative interests at the cost of restraint, coordination, and other burdens. One of the most obvious constraints necessary to treat the subjects of communicative justice as moral equals is to ensure that these benefits and burdens are fairly distributed. We'll focus first on the benefits, then on the burdens.

Attention is a scarce resource; platforms must decide how to distribute it. This raises at least two distinct distributive justice problems. One concerns the benefits of positive attention for speakers. The other concerns the benefits of the distribution of attention for audiences. Of course, these are not two distinct communities—we are all able to be both. But the distributive problems they raise are different.

Focus first on justice in the distribution of positive attention among speakers. At present, online platforms are mostly designed to maximise the allocation of attention to those who already have a lot of it. Some of this obviously comes down to consumer choice, and individuals' talent for attracting attention. Some of it derives from recommender systems aiming to optimise engagement on the platform, as well as basic features of platform design, like the dynamics of constructing platforms around people's social graph.[122] In addition to this 'Matthew Effect' (rich get richer) many online platforms also display worrying tendencies to distribute attention disproportionately to white men, and away from their complement.[123] The status quo is clearly inadequate, but what alternative principle should guide us?

Cohen and Fung defend a right to a fair opportunity for expression, but argue that this 'is not a right to have others listen'.[124] But I think we do have communicative rights to participate in relevant civic publics, and this does mean that we should at least have the opportunity to reach an audience in those publics (acknowledging that platforms can hardly force people to listen). Indeed, a basic commitment to egalitarian treatment implies we have some kind of defeasible right to be listened to, at least at a first pass, at least by some appropriate audience. We can forfeit that right through what we then say, but taking others seriously as moral equals means that at least some of us must not ignore them without cause.[125] More than this, the attention allocated by platforms can be invaluable to people, both non-instrumentally and instrumentally.[126] If a central authority is distributing some limited good, then justice is at stake, and people have a right to a just share of that good.

---

[122] Easley and Kleinberg, 2010.
[123] Geyik et al., 2019.
[124] Cohen and Fung, 2021: 29.
[125] This implies something like an imperfect duty to pay attention. A perfect duty would obviously be over-demanding.
[126] Of course, this goes only for positive attention. Negative attention can be very harmful.

I think we should take inspiration from Rawls' approach to distributive justice, and in particular his account of how justice should shape the allocation of jobs and other offices. The opportunity to reach an audience is analogous to the opportunity to hold some particular job or office: not everyone will want to take advantage of it, and its distribution should be sensitive to the talents and efforts of those who seek it. Equally, however, people should not be disbarred from accessing that opportunity on the basis of morally irrelevant factors—and it is especially objectionable if their opportunities are limited by properties that are themselves the product of systemic structural injustice. Rawls' principle of fair equality of opportunity holds up quite well as a principle to govern the distribution of (positive) attention in online communication—at least when we consider it from the perspective of the speaker.

However, the distribution of attention also has effects on the audience, determining the degree to which *their* communicative interests are satisfied. This is somewhat analogous to how Rawls thought of the distribution of income and wealth. As with Rawls' difference principle, clearly our first goal should be to ensure that people's communicative interests are satisfied as much as possible—this means, for example, allocating attention so that it optimally advances people's interests in forming accurate beliefs, in coordinating for collective action, in being entertained and so on. But we also want to ensure that these benefits are fairly distributed. In my view rather than adopt Rawls' thesis that people are treated fairly if the worst off are no worse off than they would be under any alternative distribution—a kind of primary goods prioritarianism—I think we should adopt a more structural prioritarianism. On this approach, we have strong prioritarian reasons to ensure that people in structurally marginalised or oppressed social groups do not, as a consequence of their structural position in society, enjoy worse prospects for the fulfilment of their communicative interests than those who are structurally privileged.[127] This is in part because antecedent structural disadvantage calls for redress, and part because communicative practices are otherwise likely to compound structural disadvantage. More optimistically, we also have prioritarian reasons to shape public communication to remedy structural disadvantage, for example by helping minoritized publics find themselves and work together for common goals.[128] Obviously I cannot defend this principle at any length here; the key point is that we need some such principle to address the fact that, without it, some communities' communicative interests will invariably be served much better than others'.

The analogy with Rawls' principles breaks down somewhat when we consider how they relate to one another. For Rawls, fair equality of opportunity is prior to the difference principle. First we ensure that everyone has fair equality of opportunity. Then we ensure that the resulting distribution of wealth and income maximally benefits the worst-off group. In our case, there is no obvious reason to prioritise speakers' communicative interests over those of audiences. Prima facie, we should be able to trade them off against one another—if audiences can be much better served by reducing equality of opportunity for positive attention, then that possibility should be considered. Further complexity derives from the fact that platforms only ever really distribute *potential* attention: they can surface communication so that it is visible to the user, but cannot control whether the user actually attends to what is before them. This is true with respect to both the good of positive attention for speakers, and the communicative interests served by attention of those who attend.

The same basic principles should govern distribution of the burdens of communicative justice. This takes us beyond the distribution of attention into how platforms shape public communication, especially through safety and moderation practices. Most platforms enable users to protect themselves against abuse and other forms of online harm through the ability to unilaterally block unwanted interlocutors. They therefore often face a choice between proactively enforcing egalitarian communicative norms, and leaving people to do their own unilateral policing. The latter choice is obviously distributively unfair, since it places the burden of policing harmful behaviour on the very victims of that behaviour. It is doubly unjust, given that the targets of online abuse are very often also members of structurally disadvantaged populations.[129] One reason, then, for platforms to robustly enforce communicative norms is to ensure that the burdens of promoting our communicative interests are not disproportionately borne by those with the greatest claim to avoid them.

## 5.4. Who: Authority

Suppose that the digital public sphere were governed not by platforms and governments, but by an advanced AI system. It monitors all public communication, promotes the fulfilment of people's communicative interests, and ensures both that the benefits and burdens of doing so are fairly distributed, and that people's communicative rights are protected, as they are protected against serious communicative harms by others. This AI ruler would fulfil the key demands of substantive justification. But that is clearly insufficient to know if it rules permissibly. We must also ask what gives it the right to rule the digital public sphere? On what authority does it decide what will fulfil people's communicative interests, what counts as a fair distribution, what our communicative rights are? Existing platforms and governments fall so far short of substantive justification that it is natural to focus on asking how they could clear *that* hurdle. But proper authority too is vital for the all-things-considered permissibility of governing power. Who, then, has the right to govern the digital public sphere?

Start with an obvious proposal. Perhaps platforms have the right to govern themselves (and by composition the digital public sphere) just because their users consent to their authority when they join the platform. This fact is clearly *relevant* for the justification of platform power. Those who agree to use a platform can be expected to know its rules. If those rules amount to a reasonable interpretation of how online communication should be governed, then this clearly gives platforms some authority over you. However, we should not rest on our laurels. As I argued in Lecture I, critical tech scholars have shown, time and again, the inadequacy of appeals to consent to justify the power of tech companies.[130]

First, a healthy digital public sphere is a forum for *communication*. It therefore

---

requires coordination around common protocols and architectures.[131] Individual consent is inadequate grounds for justifying the adoption of common communicative protocols—once a certain critical mass has been reached, dissent does nothing more than deprive the dissenter of access to the common network; it does not enable them to pursue some alternative public forum (just consider the tenacity of X fka Twitter as a component in the public sphere, despite everything that has been done since it went private to push people away).

Second, online platforms realise negative externalities for everyone—the spread of misinformation, harassment, and manipulation; collective and stochastic harms. Platforms are presently befouling the digital public sphere, thwarting people's communicative interests and undermining civic robustness, whether they consent to the platform's authority or not. Your consent to a given regime does little to justify the harms visited by that regime on those who do not consent.

Third, much depends here on why we care about proper authority. Are we motivated only by concern for *individual* liberty, such that your basic liberty dictates that nobody should rule you unless they have a right to do so, grounded somehow in facts about your individual choice? Or does the right to rule matter also because we care about *collective* self-determination—the idea that groups should be able to shape the shared terms of their collective existence, that they shouldn't just be passive flotsam in the chaotic maelstrom of others' free, isolated, individual choices? I think proper authority matters for both reasons—as a bulwark for individual liberty, and as a necessary means for collective self-determination.[132]

Governing the digital public sphere involves taking a stand on many controversial normative and empirical questions—both adopting an ideal of communicative justice, and taking responsibility for implementing it. It means determining what our communicative interests are, and which communicative practices, indeed which particular communicative acts, advance those interests. Many of these decisions cannot be objectively or otherwise definitively resolved. They can only be decided, so it matters immensely who decides them. The consent of some subset of participants in the digital public sphere to platform authority is too fragile a protection either for individual liberty or for collective self-determination. It is inadequate for individual liberty both because it is generally such poor quality consent, and because those who consent are not deciding for themselves alone, but also for those who dissent. It is inadequate for collective self-determination because it allows private platforms and to a lesser extent their users to determine the shared terms of groups' collective existence, and to take charge of collective, in some cases irreducibly social, goods whose steerage is vital for collective self-determination.

One might be tempted, here, to argue that platforms should endeavour to avoid *any* governing role—to pass responsibility entirely to consumers, or to governments. But this is not an acceptable response. As argued above, platforms cannot avoid shaping public communication and distributing attention, and these are the mechanisms by which the pathologies of the digital public sphere are caused, and the means by which they can be remedied (if that is even possible).

---

Non-intervention is simply not an option. And relying on consumer choice to solve coordination and collective action problems is a non-starter. Individual consumers cannot decide what networks, protocols or platforms to adopt. They cannot unilaterally determine the distribution of attention. Purely market-based approaches to communicative justice are as tendentious as purely market-based approaches to distributive justice.

Indeed, one could go further and argue that the very idea of allowing private platforms to govern the public sphere is anathema. These are private companies whose function is to generate profit. They are often run by quixotic billionaires with far too much power. Their business model is based on mass surveillance, data extraction, and arguably manipulative targeted advertising—which is itself dependent on maximising user engagement, keeping them on platform as long as possible in order to extract more data and get their eyes on more ads. One can reasonably question whether this business model could ever be consistent with advancing the goals of communicative justice. And civic robustness relies on the possibility of private communication, which is radically undermined by the surveillance practices of existing online platforms. More generally, the track record of private enforcement of public norms is poor—consider, for example, how copyright law has been enforced by for-profit online platforms, which optimise for the minimisation of liability, disproportionately prioritising the interests of copyright holders over other people.[133]

But if market-based solutions are inadequate, and if private platforms are constitutively ill-suited to governing the digital public sphere, then should we just pass platform governance entirely on to governments? Certainly democratic authorisation is the gold standard of authority, the only kind that truly ties authority to collective self-determination. And if accompanied by well-grounded basic rights, it also serves individual liberty. But we need to be careful here, on multiple grounds.

First, entrusting democratic states with control over the digital public sphere will very likely also empower authoritarians to exercise the same power. Private platforms can be an important bulwark against authoritarian state power.[134] Their ability to be so is diminished if they are routinely subordinated to state authority.

Second, as argued in Lecture I the boundaries of online platforms and of states imperfectly overlap. Our existing democratic institutions can give platforms some authority to govern, but this may still involve some people imposing their will on others who are not included in the salient democratic community. Just think, for example, of how EU regulations imposed on online platforms lead to their changing their practices worldwide, in order to reduce compliance costs. While we can perhaps make sense of Dewey's aspiration to create a 'Great Community' within a singular nation state, it is very hard to conceive of how to do so for online platforms with billions of users, from every community in the world.[135]

Still more importantly, states—whether democratic or authoritarian—are the most powerful agents in the world today and the biggest threats to communicative justice—indeed to every kind of justice (as well as its principal guarantors).

---

Counterbalancing that power is central to civic robustness. And excessive state involvement in the digital public sphere is inimical to forming independent civic publics that can hold the state to account

States obviously have some role to play in direct governance of the digital public sphere. Some communicative harms are sufficiently serious that they should be punished at law, not just as violations of terms of service. But this kind of minimal enforcement is insufficient to realise communicative justice. We cannot *exclusively* rely on states to govern the digital public sphere.

To make progress, we need to distinguish between two different problems of authority. The first concerns the source of democratic authorisation. The second concerns the nature of the authorised party. The simplest approach to the source of democratic authorisation is for democratic governments to provide a mandate for the digital public sphere, articulating a broad account of communicative justice, holding intermediaries responsible for realising that mandate. This approach—broadly consistent with intent of the Digital Services Act in the EU—should be designed so as to keep powerful states well clear of operational decisions about the governance of online communication, lest they use their power to undermine civic robustness. And what justifies also limits: platforms ought not promote values *other* than communicative justice, as doing so would exceed *their* mandate.

One could argue that even this is not sufficiently democratic, since the most active regulators (the EU and China) arguably have the least democratic legitimacy (obviously they are not equivalent, but the EU is really not very democratic). This prompts calls for innovative approaches to platform governance through new forms of digitally-enabled participatory democracy.[136] If approached seriously (not as random bot-infused polls) this could be a path worth pursuing, especially given some states' deep regulatory incapacity, as well as the mismatch between platform boundaries and territorial governments. However, one serious worry is that people really do not want to be that actively involved in self-government, and absent robust democratic institutions this will instead lead to a kind of participation theatre, in which platforms ostentatiously advertise their democratic intentions, but really have total control over the agenda as well as its implementation. Consider, for example, Meta's Oversight Board.[137] This seeks to ground authority not in democratic authorisation but in competency and independence. Within the parameters that it has been set by Meta, it has arguably delivered some good, procedurally well-made judgments. But its remit is determined by Meta, and the implementation of its judgments the same. So this does not really constitute a shift of power away from the platform. There would be a great risk of attempts at platform democracy that do not directly involve robust democratic institutions like states falling into the same trap.

Setting aside the source of democratic authorisation, a further question is what kind of entity should be its agent. Two institutional models seem feasible. On one approach, states should incorporate independent online platforms somewhat analogous to the state-funded, but independent, organisations of broadcast media like the BBC. The state can then give them a set of directives to follow ('inform, educate, and entertain'), and some degree of oversight to hold them to task, but

---

they should be resolutely independent from the states that endow them with authority.

On another approach, we rely on private platforms to be our 'digital Switzerlands', by structuring their incentives so that they serve their public function.[138] Given that attention on online platforms is driven substantially by how entertaining they are, and given that private companies tend to do a better job of creating entertaining, innovative media environments, and lastly given the push of network effects, it is unsurprising that we lack significant public options for the digital public sphere. But this just means that we face, again, the principal-agent problem of trying to induce private platforms to serve public ends. The most promising path, I think, is to change the business model of private platforms to better incentivise responsible governance of the digital public sphere. This will of course involve levying massive fines if they fail to do so. But more important still is changing the platforms' incentives so that they do not rely on engagement and surveillance to make a profit. I suspect this could be done with little actual social cost. Targeted advertising is arguably little more effective than contextual advertising. Indeed, if targeted advertising were explicitly and clearly transparent to those whose histories are being tracked and behaviour predicted, it would likely be still less effective.[139] And optimising for engagement is a rational strategy for profit-seeking platforms only because, at present, we allow them to externalise its inevitable costs. It is roughly analogous to the use of fossil fuels over renewables — economically rational only insofar as the true costs are not accounted for. The ultimate source of value here is not our data, or febrile engagement, but our attention — especially our collective attention. Optimising for engagement may not even ultimately increase long run time on platform. Even if it does, there is plenty of attention to go around without exploiting it so aggressively. This attention should be able to generate enough surplus value to sustain an independent digital public sphere, operating within a mandate articulated by democratically-elected governments, without necessitating either surveillance advertising or short-sighted extractivism.

This still leaves us with the urgent question of what we should do *now*, given the lack of adequate regulation, adequate public alternatives, and misaligned private incentives. In circumstances of institutional failure, authority can be grounded in mere competence and the lack of available alternatives. Private companies that aim to realise communicative justice can have proper authority, pro tem, on this basis. However, they (and we) should clearly endeavour to bring about regulation that can both give democratic standing to the conception of communicative justice being advanced, and shape private companies' incentives to better serve that value.

Crucially, states must not only authorise platforms to better govern online communication, they must obligate them to do so. Platforms cannot defensibly appeal to their own rights of free expression to obviate their responsibility to realise communicative justice — if A governs B, C, and D, and thereby is instrumental in shaping social relations among B, C and D, determining whether some have power over others, and materially shaping the shared terms of their

---

social existence, then A cannot say that the principles by which he governs B-D are mere acts of self-expression on his part, that the state ought have no say over.

## 5.5.  How: Legitimacy

Substantive justification and proper authority are jointly necessary for justified governance of the digital public sphere, but they are still not sufficient. As well as knowing that power is being used for the right ends, by the right people, we also need to know that it is being used in the right ways. Procedural legitimacy, as argued in Lecture I, is fundamentally about limiting power, making sure that those who govern us are confined within strict rules, so that we in turn can exercise power over them, holding them to account when they misstep or overstep. This protects individual liberty by reducing the prospects of wrongful interference, and serves collective self-determination—making those who govern us into instruments of the collective will. But it is also crucial for equality, both between rulers and ruled, and among the ruled. They have power over us, but we, by holding them to account, have power over them. And how they govern determines whether we stand in egalitarian social relations with each other: an egalitarian public sphere must be governed with an even hand.

Procedural legitimacy applies to the exercise of power in three distinct ways, between which we can distinguish temporally: ex ante, in medias res, and ex post. Before governing power is exercised, the party exercising that power should make clear the provenance of its authority, as well as make public the rules that it will govern by, so that those subject to them can comment on and ideally influence them. When ruling power is exercised, it must be applied transparently and consistently, without fear or favour. And after the fact, we should be able to audit and contest these decisions, and hold rulers to account for mistakes or abuses.

Platforms shape public communication and distribute attention. They govern the digital public sphere through their policies, to be sure. But they also do so through their architecture and algorithms. The standards of procedural legitimacy should apply not only to platforms' enforcement of their policies, but also to the technological infrastructure that they create. I will discuss 'textual' and 'technological' legitimacy in turn.

The standards of ex ante textual legitimacy are low-hanging fruit. Platforms historically fared poorly, making up policies on the fly and failing to adequately communicate them to users. Regulators, activists, journalists and scholars have, over time, done much to remedy this, though platform still rely heavily on impenetrable terms of service that almost nobody reads, and which are inconsistently applied.[140] And X, fka Twitter, has egregiously regressed against these standards since 2022.

In medias res legitimacy is both harder and more urgent for online platforms. Transparency and consistency in textual legitimacy—especially in the application of safety and moderation measures—are essential both for ensuring egalitarian relations between rulers and ruled, and for sustaining egalitarian relations among the ruled.

Principles of communicative justice grant platforms some degree of licence to

---

decide whose speech to support, of course. But if they get to secretly and arbitrarily decide whether some particular case falls within that prerogative, then it is in principle unbounded. Recognising the importance of transparency and consistency in applying platforms' policies provides users with a minimal guarantee of their moral standing with respect to the platform.

More importantly, transparency and consistency are necessary for equality before the law (or law-like rules). Communicative justice aims at fulfilling communicative interests in ways that respect our fundamental moral equality. This means enacting egalitarian relations between platform users, as well as between users and the platforms. Importantly, this applies both to platform-initiated interventions (e.g. a post is automatically flagged as being in violation of the policy) and to user-initiated interventions (e.g. a post is flagged by users as being harmful). Part of 'in medias res' legitimacy is ensuring that these kinds of disputes are fairly arbitrated, and are not systematically used by some (groups) to oppress others (as in fact is usually the case).[141]

Much of the action in discussion of procedural legitimacy in platform governance has focused on ex post criteria like contestability and accountability. Some scholars productively analogise these to individual due process rights for those subject to adverse content moderation decisions.[142] Others criticise this as 'accountability theatre', arguing that it operates at too small a scale, and that a systemic approach to platform regulation is more important.[143] This objection derives in part from a rejection of 'procedural fetishism', driven by a desire for platform regulation that focuses instead on (in my terms) substantive communicative justice. While I think substantive justification matters (and perhaps matters most), procedural legitimacy matters too. And procedural (textual) legitimacy can indeed operate at the system level as well as at the individual level.

For example, often the stakes of individual enforcement decisions will be just too low for the great effort of ensuring contestability to be proportionate. If we care about relational properties—like whether the platform is more rigorously enforcing complaints on behalf of one group as against another—then we have to audit many decisions, not give each individual the right to contest decisions that apply to them. And if we seek to review decisions made to prevent stochastic and collective harms, we cannot make *any* progress by considering individual interventions in isolation from one another.

But individual due process still matters too—at least in cases of bans and suspensions, when legitimate expectations and serious interests are at stake. And as well as being non-instrumentally important, due process can be an instrumentally valuable means to realising communicative justice. Normative regimes are most effective when they do not have to rely on perfect enforcement for their realisation, but can instead proceed through people's will, by operating on either their incentives or their values. Procedural legitimacy can support this process, by clearly informing people how their non-compliant behaviour failed to comply, and by issuing directives backed by a chain of escalating coercive threats.

For example, consider stochastic and collective harms. In these cases, by definition,

the individuals who contribute to the harm are not guilty of any especially serious act in their own right. Instead, their intrinsically insignificant speech combines with others' to produce something that is harmful in the aggregate. The best way to prevent these harms is often to filter these communications—not to take them down or penalise the author, but to simply prevent them from being seen unless explicitly searched-for. In these cases, participants in the stochastic or collective harm might have no strong claim to due process of any kind. But by notifying them of the nature of the intervention and the reasons for it, as well as by initiating some escalating system of penalties for potential future transgressions, platforms can potentially induce people to internalise these norms, so that they do not contribute to future swarms of collectively harmful behaviour.[144] Of course, this is an empirical claim. Perhaps measures like this would lead to more reactance, and more harm to innocent victims.

Platforms' moderation decisions very clearly involve the exercise of governing power, and critics can draw on well-established normative tools from the free speech literature to evaluate them. Their pre-eminence in the moral critique of platform governance is therefore understandable. I think, however, that enforcement decisions play a relatively small role in how platforms govern the digital public sphere. Platforms' architecture and amplification practices are much more central to how they shape public communication and distribute attention—to see this, note that only a small subset of content is ever a serious candidate for moderation, whereas *everything* is modified by platform architecture and subject to curation. However, we lack a well-worked out theory of how to apply norms of procedural legitimacy to this kind of platform governance. I won't attempt to supply such a theory here, but will instead indicate how communicative justice requires at least some specific standards to obtain.

Ex ante, platforms' design choices should be explicit and open to some degree of review; in particular, it should be clear what alternative paths could have been adopted. As an illustration of this, consider the implementation of platform governance in one of the newest platforms for online communication: ChatGPT, and the broader OpenAI API. OpenAI's LLMs contribute to and will in future shape public communication in significant ways. And each is subject to a vast suite of rules conceived by OpenAI as means of 'aligning' the language model, but in reality operating as mechanisms of governing their users, limiting in opaque ways how they can deploy OpenAI's models to advance their communicative goals. OpenAI has specific user terms and conditions, policies that constrain how users can communicate using their models. But those policies are not implemented in the models in any kind of robust or simple way, and the models are prone to baulk at uses that are in fact compliant with those policies. Legitimate algorithmic governance requires committing ex ante to a set of *public* rules, standards and ideals—within the scope of a democratic authorisation—and then governing on that basis. OpenAI—and other companies shipping dialogue agents based on large language models—fall well short of that mark.

In medias res, we need to balance concern for the individual and systemic impacts of platforms' architecture and algorithms. In particular, it is not enough to ensure that individuals are treated in a way that is transparent and consistent; we must insist on insight into the systemic effects of those individual choices. This is because, first, *these are systemic choices*. Platform design—architecture, moderation and curation—affects everyone (if it doesn't, then that's a different issue). So considering their legitimacy from the individual perspective is rather like asking whether the constitution of the labour market is legitimate from an individual's perspective. Second, collective and relational values are at stake. Civic robustness and a healthy information environment, and perhaps a vibrant creative economy, are collective goods; they may also be irreducibly social goods. And the fair distribution of the benefits and burdens of communicative justice is a relational value. Due to these collective goods, we should be sceptical of attempts to derive procedural legitimacy for platform governance through the introduction of greater consumer choice.

---

[144] In addition, if filtering is done intransparently, then people will be prone to infer that they are being 'shadowbanned' when they are not, which can in the end undermine the legitimacy and authority of the ruler ([West, 2018](#)).

This insight has broad application. For example, recommender systems should aim not at local but at global optimisation. Rather than addressing the question, 'for this user at this time, what content would be most relevant?', they must at least also ask 'for this user at this time, in this community, what content would best contribute to realising communicative justice in this community?'. And we should be sceptical of the promise of 'middleware', i.e. recommender systems that can be plugged into existing social media platforms to provide users with greater choice over what they see online.[145] First, this approach reifies the role of recommender systems in isolation from the broader sociotechnical systems of which they are part. More importantly, the digital public sphere's pathologies are caused in part by aggregated locally optimal choices leading to globally suboptimal outcomes.[146] We are unlikely to realise collective and relational goods by doubling down on the individualist approach. Reliance on competition law or antitrust approaches to fixing the digital public sphere face similar objections.[147] Undoubtedly they have a role to play. But if many of the goods at stake are collective, then there are returns to scale for communicative justice, and empowering consumers risks exacerbating the problem, not fixing it.

Ex post legitimacy in platform design should similarly focus on systemic questions. Our target is not to ensure that individuals are able to contest decisions to amplify or ignore their posts, more to see to it that the whole system is subject to appropriate oversight, with systemic impacts being contestable on behalf of the relevant political community as a whole (and with accountability for bad systemic choices). This is where demands for platform observability—with respect to data, models, and algorithms as well as human decision procedures—should sit.[148] We cannot effectively hold platforms accountable for the pathologies of the digital public sphere, or task them with advancing communicative justice, if we lack access to the data that would demonstrate the impact they are (or are not) having, as well as the technological artefacts that determine that impact. Calls for platform transparency and algorithmic accountability are, on this view, calls for ex post technical procedural legitimacy in the pursuit of communicative justice.

The advent of extremely capable LLMs offers a potentially new spin on the idea of

---

middleware, which it is worth pausing to explore—as it might offer the most concrete and promising prospect yet of changing the underlying structures that reliably lead us away from communicative justice in the digital public sphere.[149] LLMs like GPT-4o could potentially underpin automated attention allocators that are much better able to understand the content of what we might view online, and to capture not imperfect proxies, but the essence of our preferences, both higher- and lower-order. Besides their impressive ability to functionally understand natural language, as mentioned in Lecture I, these models can also be fine-tuned to function as agents, using tools (other software through APIs) to perform a wide range of different functions. We could develop a Language Model Agent partly inspired by the Conversational Recommender Systems that already exist, the goal of which would be to elicit user preferences using natural language, rather than observing behaviour, and then to browse the internet and the user's social media feeds identifying content that serves those preferences.[150] Rather than relying on the weak proxy of *engagement* to infer our preferences, the Agent could directly ask us what we want to see, and when it shows us something we do not want to see could ask us why.

GPT-4o is already incredibly capable at understanding moral language and user preferences, as well as at discerning whether some particular piece of content is likely to match those preferences. While language itself can sometimes be an inefficient user interface, and would clearly not be the be all and end all of the user experience, it would enable a Language Model Agent to fill in the unarticulated gaps in our preferences in ways that we rationally endorse. This might amount to being able to recognise what's a post by an AI influencer, and what's a post of genuine AI research. Or it might mean picking up spoilers for *any* show, not just ones you've had the foresight to mute. Or it might mean helping you find a range of ideological perspectives on some issue of the moment.

Language Model Agents like this—algorithmic intermediaries in the truest sense—could also help us guide our online behaviour in our more considered moments, rather than just profiting from our tendency to slip into automatic behaviour, as existing recommender systems do. We could talk with our Agents about how we want to use the internet, and what kinds of behaviours we later regret. The Agent could respectfully nudge us when we're slipping, reminding us of what we earlier said we wanted to do—for example, catching us as we are about to post an angry reply online, or else reminding us when we're doomscrolling to get outside and touch grass. We could also design Language Model Agents to take into account and try to mitigate the kinds of collective harms discussed above; even a degree of coordination would be possible between Agents. Better still, these agents would be able to involve us in the practice of allocating our own attention—explaining why they 'thought' a particular post would be of interest using natural language in a veridical way.[151]

Perhaps most exciting, these Language Model Agents could in principle upend the political economy of attention itself. As noted above, many of the pathologies of the digital public sphere derive from the basic business model of online platforms, and in particular the need to gather vast quantities of user data in order to indirectly infer peoples' preferences, predict what they will find engaging, and serve it to them. This is the intersection of surveillance capitalism and engagement-based optimisation with which I opened this book. Language Model Agents can in principle understand the content and context of a text, image or video post, and elicit and round out a user's preferences directly rather than through revealed behaviour, and then match the content to the preferences. This could enable an alternative model for the allocation of attention that does not rely on surveilling everyone's behaviour online. And it need not rely on optimising for engagement (though of course it could). Since Language Model Agents like these would browse the internet on our behalf and identify content for us, they could in principle be designed to operate just in our interests—not to hold our attention and serve us ads. In fact, while at present only the most advanced, largest models would be able to function effectively in this role, computational and algorithmic advances suggest that it may before long be possible to run such an agent locally on a smartphone, or in a private cloud. This could definitively break the large online platforms' stranglehold on attention. There is no reason why many different Language Model Agents could not be developed—nothing necessitates or even implies that such systems would lend themselves to a monopoly.

---

[151] LLMs are not in general good at explaining themselves, but in this case the Agent would have a natural language description of your values, and would be selecting posts that it construed to optimise for those values, so it would be able to offer a factive explanation.

Obviously relying on a Language Model Agent to serve as one's intermediary to the digital public sphere would be risky, and the underlying idea is undoubtedly more likely to be acted on by the very platforms that such Agents could supplant. Or else, the next generation of digital capitalism might just end up giving the proprietors of the underlying models the kind of power that Google, Meta and TikTok currently have. And there may remain some fundamental obstacles to such Agents' successful operation—perhaps even with a deep understanding of a user's preferences, and of the content of posts on the internet, lacking vast amounts of data from millions of users would be too much of a limitation. But, as argued above, the pathologies of public communication have in many ways derived from the affordances of the algorithmic tools at our disposal when constructing the digital public sphere: in particular, recommender systems based on big data analytics, and latterly deep reinforcement learning.

These systems have three big problems. They rely on mass online surveillance. They are in the private control of profit-seeking corporations that are ultimately optimising for their profits. And when they allocate our attention, they have only imperfect proxies from which to infer both the nature of the content that they are serving, and the value it has for those to whom it is served. Language Model Agents would have different affordances. They would not depend on mass surveillance, they could be optimised for their user's interests not for some private corporation, and they could functionally understand both communications online and their user's true preferences. This would not solve all our problems—as I have acknowledged throughout, our failure to realise communicative justice is as much due to us as to the algorithmic intermediaries by which we are connected. But it would give us new levers to tilt the affordances of those intermediaries in the favour of communicative justice. And if these systems could genuinely be in our own control, they could actually support and scaffold a kind of algorithmic *self*-governance, of a truly novel kind.

## 6. CONCLUSION

My argument began with the assumption that the digital public sphere is in poor health, afflicted in particular by epistemic pollution, abuse, and manipulation. I argued that algorithmic intermediaries govern the digital public sphere by shaping communication and distributing attention. Their architecture, and moderation and curation practices, are very likely implicated in the pathologies of the digital public sphere. Even if they are not, they are our most promising levers for fixing it. They govern the digital public sphere, and as such are instrumental to shaping power relations between us, and risk unilaterally shaping the shared terms of our social existence. They are therefore obliged—and should be compelled—to do better.

If we are going to better govern the digital public sphere, it is not enough to name and individually target its pathologies. We need a positive ideal to aim at, something to thread our different goals together, help guide us in making trade-offs, and show us how the nature of our ends constrains the means by which they may be permissibly pursued. The most obvious resource in political philosophy, the literatures on freedom of expression and the democratic public sphere, can provide useful insights into the interests served by public communication, but offer a restrictive normative palette, too focused on protecting individual expression against state intervention, and ill-adapted for a question—how to shape communication and distribute attention—for which non-intervention is not a feasible answer. We need a theory of communicative justice as well—a moral theory guiding us specifically in the task of shaping public communication and distributing attention.

Theories of justice are called for when the joint pursuit of a common good by moral equals requires some to make sacrifices or show restraint for the sake of the common good, and yields benefits and burdens that can be variously distributed. A theory of communicative justice requires an account of its currency—the good that communicative justice aims to promote—and of the norms constraining that pursuit, which reflect the fundamental moral equality of those in that community. My first hope is that this Lecture has made the case for a theory of communicative justice. My second hope is to have developed that theory in an attractive, though necessarily incomplete, direction.

Drawing on existing work in political philosophy, I argued for non-instrumental and instrumental individual communicative interests in recognition, esteem, dialogue, equal treatment, knowledge, entertainment (and a vibrant creative economy), coordination and community, as well as collective interests in civic robustness, and in a healthy information environment. I then applied the theory developed in Lecture I to build an account of the norms constraining the promotion of those goods—focusing in turn on substantive justification, proper authority, and procedural legitimacy.

On the first, I argued for the importance of at least striving for pluralism in defining the currency of communicative justice, and for an account of the baselines for admissible communication in the digital public sphere that draws inspiration from theories of freedom of expression without simply duplicating them, and made a first pass at understanding the distributive questions raised by communicative justice. On proper authority, I argued for the primacy of democratic authorisation, but also the need for robust protections for an independent public sphere that is not beholden to any particular government—of special importance too given the transnational nature of the digital public sphere. And I argued that both platforms' policies and their design should be subject to the requirements of procedural legitimacy; sometimes these should be patterned on individual rights to due process with respect to the state, but sometimes they should be more focused on systemic properties of these platforms, their net effects on the digital public sphere.

This Lecture responds to the call to action given in Lecture I. I argued there that algorithmic intermediaries govern the social relations that they mediate, and that their governing power must meet standards of substantive justification, proper authority, and procedural legitimacy. In this Lecture, I have shown how a particular kind of algorithmic intermediary in a particular corner of the Algorithmic City—online platforms in the digital public sphere—govern online communication. And I have outlined a theory of communicative justice as an account of how to answer the what, who, and how questions. I have not attempted to give a comprehensive theory of communicative justice. There are many further questions to address, and I think more work should be done to limn the contours between communicative justice and other justice domains.[152] The value of communicative justice also clearly implicates many other institutions and practices besides just platform governance of online communication. In addition, I cannot here apply my high level theory of communicative justice to practical, concrete cases. Many thorny details will be resolved by bringing ideals and reality closer together. In the manner of Rawls' reflective equilibrium, our understanding of the ideals will evolve as they are tested against one hard case after another. But the guiding theme of this Lecture, and its central thesis, is that the power to shape public communication and to distribute attention demands the attention of political philosophers.[153] It is not unique to the Algorithmic City, but the scale of the impact of these practices of constituting, moderating and curating public communication is greater than ever before, and our collective ability to shape communication and distribute attention is more effective, and more fine-grained, than ever before.

In an influential essay, Nancy Fraser argued that in the presence of significant background inequality, equality in the public sphere is an unattainable goal.[154] Similar concerns are pervasive in reflection on technology and society. In one domain after another, we confront the apparent futility of aiming at local justice in the presence of radical background injustice.[155] More than futility, aiming at local justice might sometimes even legitimate background inequality—just as trying to achieve a fair distribution of labour among enslaved people would implicitly condone the fact that *they are enslaved*. The presence of extreme background inequality undoubtedly makes achieving communicative justice much harder. And the degree of inequality and affective polarisation in countries like the United

---

[152] For example, I think that some concepts currently parked in other sites of justice fit better here—for example, I think that much of testimonial injustice is rather a matter of communicative justice than epistemic justice. See Hill Collins, 1990; Williams, 1991; Fricker, 2007.

[153] I also think that while communicative justice rightly considers the distribution of attention through the lens of political philosophy, we need also to ask questions about the *ethics* of attention, and in particular how I should allocate my own attention.

[154] Fraser, 1990: 77. She was drawing on earlier work by Jane Mansbridge.

[155] See e.g. Davis et al., 2021.

States might make any aspirations to realise communicative justice seem naively utopian.

And yet perhaps we have grounds for hope. Responding to Fraser, Iris Marion Young argued that a healthy public sphere can help us bootstrap our way out of inequality.[156] Civic robustness enables power to be held accountable, and lays the discursive foundations of collective action. Only by forming robust civic publics will we change our background social structures. Achieving a measure of communicative justice within some domain is more tractable than trying to fix the whole of society in one stroke. Civic robustness is not the same as democracy. Fixing democracy might be beyond us; realising civic robustness could be more tractable. And tangible improvements can snowball. Background inequality is undoubtedly a drag on the realisation of communicative justice, but approaching communicative justice is likely to be our most effective and promising means for remedying background inequality, as well as for tackling the deep moral disagreement that underpins broader political dysfunction.

This is, I think, the most important lesson to take from thinking about communicative justice in the digital public sphere. Whatever you hold social media companies responsible for, the digital public sphere could surely be a powerful engine of positive social change, even if it presently falls far short of that aspiration. Of course, potential is not actuality. Architecture, moderation, and curation of communication can shape behaviour, but they cannot determine it. The ultimate cause of the pathologies of the digital public sphere is just people. And out of our crooked timber, perhaps no algorithmic intermediaries will make anything straight. But we should design our digital public sphere to at least express our commitment to communicative justice, even if that is insufficient for realising broader social goals. If our divisions are ultimately too great for communicative justice to form a bridge between us, then so be it. We cannot escape the obligation to try.

---

[156] Young, 2000: 50.

## Acknowledgements

I gave lectures based on these essays in Stanford, in January 2023, as two parts of a Tanner Lecture in AI and Human Values. My deep thanks go to my host, Rob Reich, and to the organising institutions, the Institute for Human-Centered AI and the McCoy Family Center for Ethics in Society, as well as of course the Tanner Foundation, in particular Beth James. I was incalculably fortunate in my commentators, whose work has much shaped my own. Thanks to Joshua Cohen, Marion Fourcade, Renee Jorgensen, and Arvind Narayanan.

Drafts of Lecture I were generously read by the members of the Machine Intelligence and Normative Theory (MINT) lab at ANU, as well as Danielle Allen, Garrett Cullity, Marcello di Bello, Gideon Futerman, Rob Gorwa, James Grimmelmann, Massimo Renzo, Nic Southwood, Charlotte Unruh, and Meg Young. Many thanks to them all for providing such invaluable feedback. I presented versions at Harvard (as the 2022 Mala and Solomon Kamm Lecture in Ethics), and then at Carnegie Mellon University, Emory University, Princeton's University Center for Human Values, and the Digital Life Initiative at Cornell Tech; thank you to my hosts and my audiences for the opportunity and the penetrating feedback.

I presented Lecture II in Geneva at the Graduate Conference in Political Theory, in Melbourne to the ADM+S Centre of Excellence, at the Knight First Amendment Institute, and to the Cornell Ethics and Public Affairs colloquium. My thanks to my hosts in each of those cases, and in particular to Jonathan Stray for his discussion of the paper in Columbia. For reading drafts of earlier versions, my thanks to Mark Andrejevic, Christian Barry, Étienne Brown, Agnes Callard, Jennifer Davis, Anne Gelling, Brian Hedden, Jeffrey Howard, Regina Rini and the members of the Machine Intelligence and Normative Theory (MINT) Lab, ANU.

For reading and providing feedback on both lectures, particular thanks to Henry Farrell, Niko Kolodny, Leif Wenar, Kyle van Oosterum, Henrik Kugelberg, Lorenzo Manuali, Shang Long Yeo and Elad Uzan.

My research for these Lectures has been funded by an Australian Research Council (ARC) Future Fellowship Award (FT210100724), an ARC Linkage Award (LP210200818), and a Templeton World Charity Foundation Diverse Intelligences project grant.

Lastly, a special thanks to Lu, Moss, and Ash, for infinite cups of tea, unstinting support, and comic relief (Lu), long conversations about AI as we walk up Mount Livingstone (Moss), and daily reminders of the intrinsic value of asserting one's autonomy in the face of governing power (Ash).